\documentclass{article}
\usepackage[utf8]{inputenc}
\usepackage{graphicx} 
\usepackage{natbib}
\usepackage{amsmath}
\usepackage{amssymb}
\usepackage{exscale}
\usepackage[affil-it]{authblk} 
\usepackage[ruled,linesnumbered]{algorithm2e}
\usepackage{algpseudocode}
\usepackage{listings}
\usepackage{float}
\usepackage{mathtools}
\usepackage{relsize}
\usepackage{xcolor}
\usepackage{pgf, tikz}
\usetikzlibrary{arrows, automata}
\usepackage{bm}
\usepackage{mdframed}
\usepackage[ampersand]{easylist}
\usepackage{enumitem}
\usepackage{tikz}
\usepackage{tkz-euclide}
\usepackage{authblk}
\usepackage{natbib}
\usepackage{multirow}
\usepackage{xr-hyper}
\usepackage[hyperfootnotes=false]{hyperref}
\usepackage{geometry}
\usepackage{array}
\usepackage{booktabs}
\usepackage{titlesec}
\usepackage{tabularx} 
\usepackage{rotating}
\usepackage{makecell}
\usepackage{indentfirst}
\usepackage{mdframed}  
\usepackage{amsmath,amssymb,amsthm} 
\usepackage{longtable}
\usepackage{booktabs}

\newtheorem{assumption}{Assumption}

\newtheorem{assumptionalt}{Assumption}[assumption]

\newenvironment{assumptionp}[1]{
  \renewcommand\theassumptionalt{#1}
  \assumptionalt
}{\endassumptionalt}

\newtheorem{conclusion}{Conclusion}

\newtheorem{remark}{Remark}

\usetikzlibrary{shapes,decorations,arrows,calc,arrows.meta,fit,positioning}
\tikzset{
	-Latex,auto,node distance =1 cm and 1 cm,semithick,
	state/.style ={ellipse, draw, minimum width = 0.7 cm},
	point/.style = {circle, draw, inner sep=0.04cm,fill,node contents={}},
	bidirected/.style={Latex-Latex,dashed},
	el/.style = {inner sep=2pt, align=left, sloped}
}

\hypersetup{colorlinks=true,allcolors=[rgb]{0,0,1}}

\newcommand\independent{\protect\mathpalette{\protect\independenT}{\perp}}
\def\independenT#1#2{\mathrel{\rlap{$#1#2$}\mkern2mu{#1#2}}}

\title{Mendelian Randomization Methods for Causal Inference: Estimands, Identification and Inference}
\author{Minhao Yao$^1$, Anqi Wang$^2$, Xihao Li$^{3,4}$, Zhonghua Liu$^{5*}$}
\date{\small $^1$ Department of Statistics and Actuarial Science, The University of Hong Kong, Hong Kong \\
$^2$  Department of Neurology, Columbia University Irving Medical Center, New York, NY, USA \\
$^3$ Department of Biostatistics, University of North Carolina at Chapel Hill, Chapel Hill, NC, USA \\
$^4$ Department of Genetics, University of North Carolina at Chapel Hill, Chapel Hill, NC, USA \\
$^5$ Department of Biostatistics, Columbia University, New York, NY, USA
\\ 
$^*$ Correspondence to: Zhonghua Liu (\url{zl2509@cumc.columbia.edu})}

\begin{document}

\maketitle

\begin{abstract}
Mendelian randomization (MR) has become an essential tool for causal inference in biomedical and public health research. By using genetic variants as instrumental variables, MR helps address unmeasured confounding and reverse causation, offering a quasi-experimental framework to evaluate causal effects of modifiable exposures on health outcomes. Despite its promise, MR faces substantial methodological challenges, including invalid instruments, weak instrument bias, and design complexities across different data structures. In this tutorial review, we provide a comprehensive overview of MR methods for causal inference, emphasizing clarity of causal interpretation, study design comparisons, availability of software tools, and practical guidance for applied scientists. We organize the review around causal estimands, ensuring that analyses are anchored to well-defined causal questions. We discuss the problems of invalid and weak instruments, comparing available strategies for their detection and correction. We integrate discussions of population-based versus family-based MR designs, analyses based on individual-level versus summary-level data, and one-sample versus two-sample MR designs, highlighting their relative advantages and limitations. We also summarize recent methodological advances and software developments that extend MR to settings with many weak or invalid instruments and to modern high-dimensional omics data. Real-data applications, including UK Biobank and Alzheimer’s disease proteomics studies, illustrate the use of these methods in practice. This review aims to serve as a tutorial-style reference for both methodologists and applied scientists.


\end{abstract}

\section{Introduction}

\subsection{Motivation for causal inference in observational studies}
A central goal of causal inference is to assess the causal relationships between variables \citep{holland1986statistics, pearl2009causal, miguel2023causal, imbens2024causal}. This involves determining whether changes in one variable (the treatment or exposure) directly influence changes in another variable (the outcome). Randomized experiments are generally regarded as the gold standard study design  in statistical research and practice due to their ability to facilitate  causal inference \citep{neyman1923applications, fisher1935design}. In essence, these experiments employ random assignment to allocate participants to treatment and control groups, which ensures that the comparison groups are balanced regarding all (measured and unmeasured) covariates except for the treatment assignment itself \citep{neyman1923applications, fisher1935design, rubin1977assignment, imbens2015causal}. This randomization minimizes bias and enhances the internal validity of the findings. In contrast, observational (non-randomized) studies are often employed when randomization is unfeasible or unethical \citep{rubin2007design,rosenbaum2010design}. Observational studies aim to draw causal conclusions from real world data; however, such studies can be affected by confounding variables—factors that may influence both the treatment assignment and outcome variables, complicating establishing treatment-outcome causal relationships \citep{miguel2023causal}.

\subsection{Mendelian randomization as a natural experiment}

An instrumental variable (IV) serves as a powerful tool in causal inference by leveraging a natural experiment, allowing researchers to uncover causal relationships even in the presence of unobserved confounding \citep{wright1928tariff,angrist1996identification, angrist2009mostly, baiocchi2014instrumental, wooldridge2016introductory}. By leveraging an exogenous source of variation, such as genotype \citep{katan1986apoupoprotein, davey2003mendelian, vanderweele2014methodological} or draft lottery \citep{angrist1990lifetime}, the IV approach isolates the variation in the treatment variable that is as good as randomly assigned, much like a randomized controlled trial. This natural experiment framework helps address confounding concerns, providing more credible estimates of causal effects \citep{dunning2012natural}.  Embracing IV as a natural experiment not only strengthens empirical research but also brings us closer to the gold standard of causal inference using randomized experiments.

\begin{figure}[!htb]
    \centering
    \includegraphics[width=1\linewidth]{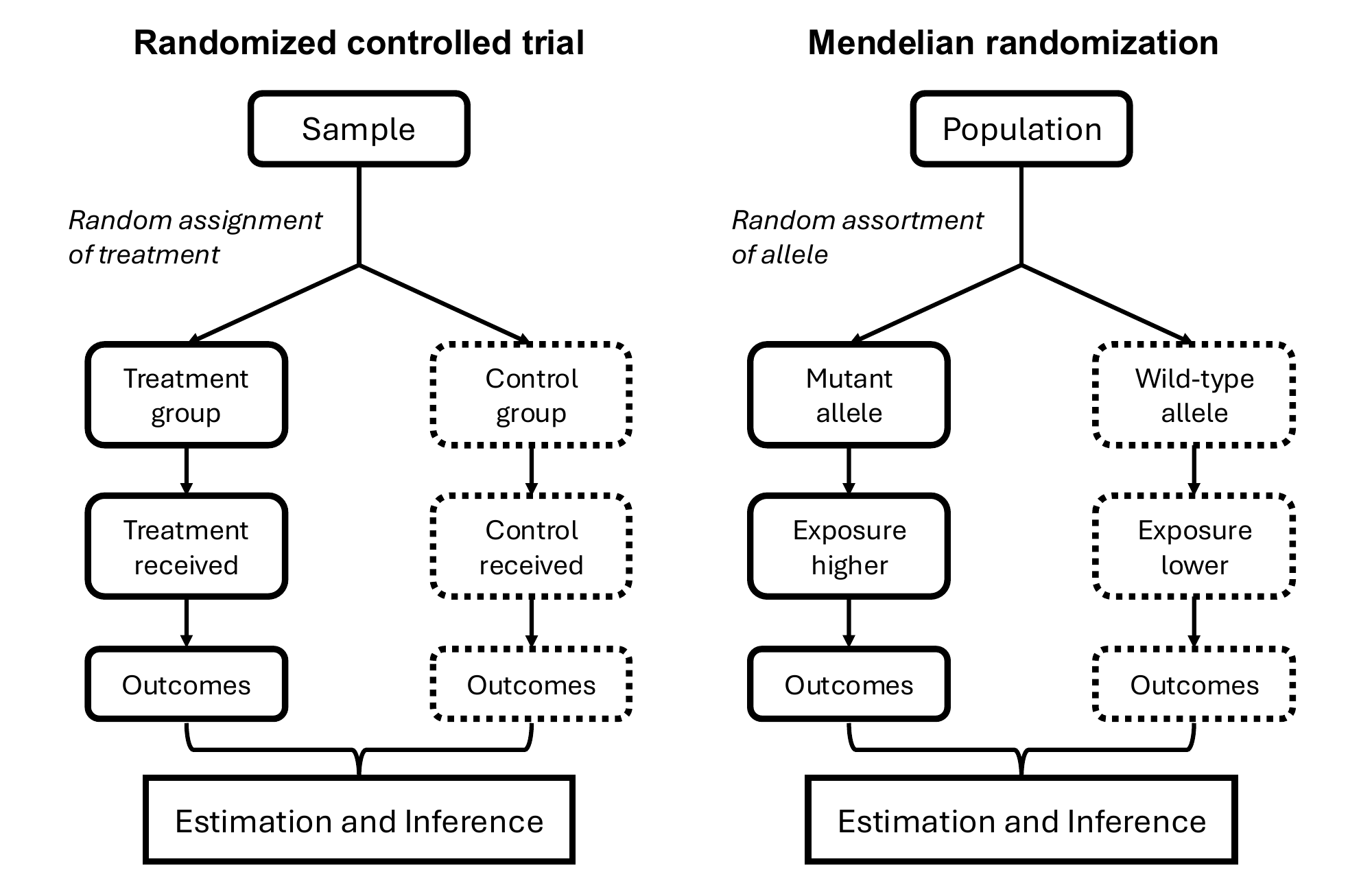}
    \caption{Comparisons of randomized controlled trial and Mendelian randomization.} 
    \label{fig: rct & mr}
\end{figure}

Mendelian randomization (MR) is a causal inference method that applies Mendel’s laws of inheritance, using genetic variants as instrumental variables to assess causal relationships between modifiable risk factors and health outcomes. By leveraging Gregor Mendel’s principles of random segregation and independent assortment of alleles, MR mimics a randomized experiment, reducing confounding biases inherent in observational studies \citep{davey2003mendelian, lawlor2008mendelian, davey2014mendelian, sanderson2022mendelian}. For illustrative purposes, we compare the designs of a randomized experiment and Mendelian randomization in Figure \ref{fig: rct & mr}. Since genetic variants are randomly assigned at conception, much like the randomization in a clinical trial, they serve as ideal instruments to assess causal relationships between modifiable exposures (e.g., cholesterol levels) and health outcomes (e.g., heart disease) \citep{thanassoulis2009mendelian, palmer2012using, emdin2017mendelian}. By leveraging the unconfounded nature of genetic inheritance, MR minimizes biases from reverse causation and unmeasured confounding, offering a robust framework for causal inference in biomedical research \citep{lawlor2008mendelian, burgess2021mendelian,sanderson2022mendelian}. This approach has been transformative in public health and medicine, helping to validate drug targets, debunk spurious associations, and guide public health policies \citep{haycock2016best,smith2017will,yao2024deciphering}. By employing genetic variants as IVs, MR leverages the natural randomization of alleles conferred by Mendelian inheritance, transforming observational data into a quasi-experimental framework that robustly infers causal relationships.

\subsection{From causal estimands to statistical inference}

In this paper, we adopt the \textit{estimand framework} to elucidate key concepts, study designs, statistical inference methods, and causal interpretations in causal inference, aiming to clarify common misconceptions and provide practical guidance for MR analysis \citep{lundberg2021your,kahan2024estimands}. By formally defining causal estimands, such as the average treatment effect or local average treatment effect, we align MR with the underlying causal questions of interest \citep{angrist1995identification, imbens2004nonparametric, lundberg2021your}. We then discuss how MR designs and analytical approaches target these causal estimands under certain assumptions \citep{haycock2016best, ference2021using}. This estimand framework  not only enhances the interpretability of MR results but also helps researchers navigate methodological challenges, such as pleiotropy and weak instrument bias, ensuring more reliable causal inference in practice \citep{lewis1999statistical, vanderweele2016commentary, little2021estimands, han2023defining, keene2023estimands, kahan2024estimands}. 

\begin{figure}[!htb]
    \centering
    \includegraphics[width=\linewidth]{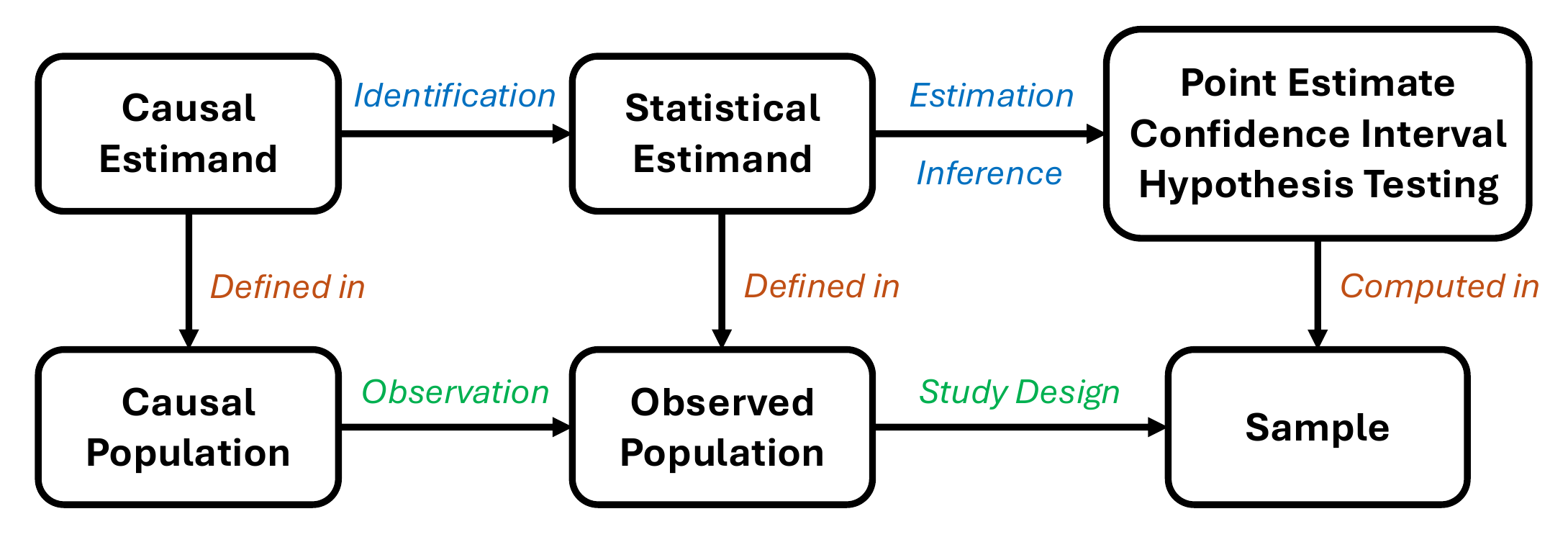}
    \caption{Conceptual flowchart bridging causal population, observed population, and sample in causal inference.}
    \label{fig: causal flowchart}
\end{figure}

To formulate a coherent causal inference framework, it is essential to distinguish the following three key concepts: causal population, observed population, and sample, as illustrated in Figure \ref{fig: causal flowchart}. The \textit{causal population} consists of all subjects in the study domain, where each subject is associated with multiple potential outcomes, one corresponding to each level of the treatment or exposure \citep{neyman1923applications, rubin1974estimating, angrist1995identification, angrist1996identification, rubin2005causal}. Causal estimands are precisely defined target quantities in causal population specifying the causal effects that we are interested in. The \textit{observed population} consists of subjects for whom only one potential outcome is realized due to the actual treatment assignment. Statistical estimands are quantities that are defined in the observed population. Causal assumptions (e.g., consistency, unconfoundedness) are required to establish causal identification, linking a causal estimand to its corresponding statistical estimand defined in the observed population \citep{neyman1923applications, rubin1974estimating, han2023defining}. A \textit{sample} consists of a subset drawn from the target observed population through either random or non-random selection procedures during data collection. Estimation and inference within this sample necessitates accounting for the sampling design to draw valid conclusions about the target observed population.

\subsection{Outline of the paper}
The remainder of the paper is organized as follows. Section~\ref{sec: natural experiments} introduces how natural experiments based on genetic variation underpin MR analysis. Section~\ref{sec: estimand} defines key causal estimands and the identification assumptions under the potential outcomes framework. Section~\ref{sec: ALICE} focuses on causal estimand under the Additive LInear Constant Effect (ALICE) model framework. Section~\ref{sec: invalid IV} discusses identification and inference when some genetic instruments are invalid. Section~\ref{sec: weak IV} introduces weak instrument bias and reviews recent methods aiming at mitigating this bias. Section~\ref{sec: family-based} compares MR analyses based on population-based versus family-based study designs. Section~\ref{sec: ind vs sum} discusses the use of individual-level and summary-level data in MR, highlighting their respective advantages and limitations. Section~\ref{sec: one-sample vs two-sample} compares one-sample and two-sample MR designs, focusing on the assumptions and the behavior of weak instrument bias. Section~\ref{sec: IV selection} outlines the procedure and key considerations for selecting genetic instruments in MR analyses. Section~\ref{sec: real data} illustrates method comparisons using two real-data applications. Finally, Section~\ref{sec: future} outlines future directions for MR, including binary and survival outcomes, longitudinal designs, and multivariate MR.

\section{Using Genetic Variants as Instruments: Natural Experiments in Health Research}
\label{sec: natural experiments}
\subsection{From randomized trials to natural experiments}
Randomized controlled trials (RCTs) serve as the gold standard to establish the causal relationship between an exposure and an outcome \citep{fisher1935design, stolberg2004randomized}. However, some exposures are unethical or even infeasible to be randomized \citep{hellman1991mice,goldstein2018ethical}. As an alternative to RCTs, natural experiments are observational (non-randomized) studies where subjects are assigned to the treatment or control groups based on events determined by other factors beyond the control of researchers \citep{dinardo2010natural, dunning2012natural, craig2017natural}. Natural experiments are common and have been used extensively in many fields, especially when the exposure in view cannot be ethically or practically manipulated in experimental settings \citep{sanson2014evaluation, craig2017natural, leatherdale2019natural}.

\subsection{Genetic inheritance as a natural experiment}

Mendelian randomization (MR), named after Gregor Mendel (1822–1884) who established the laws of  Mendelian inheritance \citep{castle1903mendel,biffen1905mendel, bateson2013mendel}, leverages the random assortment of genetic information during meiosis as a natural experiment to assess the causality between a modifiable exposure and an outcome of interest from observational studies \citep{davey2003mendelian, lawlor2008mendelian, davey2014mendelian}.  In biallelic single-nucleotide polymorphisms (SNPs) where two possible alleles exist at a specific locus, the predominant allele in the population is referred to as the wild-type or major allele, while the less common allele is referred to as the variant or minor allele \citep{international2005haplotype, chari2013conditional}. During meiosis, alleles for unlinked genes are inherited independently, which is a process governed by Mendel's Law of Independent Assortment \citep{castle1903mendel,biffen1905mendel, kleckner1996meiosis}. This process forms the basis for the natural experiment underpinning MR framework \citep{davey2003mendelian,lawlor2008mendelian,davey2020mendel}.

\subsection{Instrumental variable assumptions and potential violations in MR}

Just as Archimedes famously claimed, ``Give me a place to stand, and I will move the Earth'' \citep{dijksterhuis2014archimedes}, IV methods echo: ``Give me a valid instrument, and I will eliminate confounding.''  For reliable causal findings, genetic instruments included in the conventional MR analysis are required to be valid IVs, that is, they should satisfy the following three core IV assumptions \citep{lawlor2008mendelian, didelez2007mendelian}:
\begin{assumptionp}{A1}[IV relevance]
    The genetic variant is associated with the exposure. \label{ass: IV relevance}
\end{assumptionp}
\begin{assumptionp}{A2}[IV independence]
    The genetic variant is not associated with unmeasured confounder of the exposure-outcome relationship. \label{ass: IV independence}
\end{assumptionp}
\begin{assumptionp}{A3}[Exclusion restriction]
    The genetic variant affects the outcome only through the exposure. \label{ass: exclusion restriction}
\end{assumptionp}

\begin{figure}[!thp]
	\centering
	\begin{minipage}{0.49\textwidth}
		\centering
		\begin{tikzpicture}
		\node[state] (1) {$Z$};
		\node[state] (2) [right =of 1] {$D$};
		\node[state] (3) [right =of 2] {$Y$};
		\node[state] (5) [above =of 2,xshift=.9cm, yshift=-0.2cm] {$U$};
		\path (1) edge[->,-Triangle] node[above] {} (2);
		\path (2) edge[->,-Triangle] node[above] {} (3);
		\path (5) edge [->,-Triangle]node[el,above] {} (2);
		\path (5) edge [->,-Triangle]node[el,above] {} (3);
		\end{tikzpicture} \\ \vspace{6mm}(a) valid IV
	\end{minipage}\hfill
	\begin{minipage}{0.49\textwidth}
		\centering
		\begin{tikzpicture}
		\node[state] (1) {$Z$};
		\node[state] (2) [right =of 1] {$D$};
		\node[state] (3) [right =of 2] {$Y$};
		\node[state] (5) [above =of 2,xshift=.9cm, yshift=-0.2cm] {$U$};
		\path (1) edge[->, thick, color = red,-Triangle,dashed] node[above] {} (2);
		\path (2) edge[->, -Triangle] node[above] {} (3);
		\path (5) edge[->, -Triangle] node[el,above] {} (2);
		\path (5) edge[->, -Triangle ] node[el,above] {} (3);
		\path (5) edge[->, thick, color=green, -Triangle] node[el,above] {} (1);
		\path (1) edge[->, thick, color = blue,-Triangle,bend right=40] node[el,above] {} (3);
		\end{tikzpicture}\\(b) Invalid IV
	\end{minipage}
	\caption{Directed acyclic graphs (DAGs) that show the relationship among an instrumental variable $Z$, a treatment/exposure $D$, an outcome $Y$, and the unmeasured confounding $U$. In the right DAG, the dashed red, solid green and blue lines represent violations of the IV relevance (\ref{ass: IV relevance}), IV independence (\ref{ass: IV independence}) and exclusion restriction (\ref{ass: exclusion restriction}) assumptions, respectively. }
	\label{fig:dag}
\end{figure}
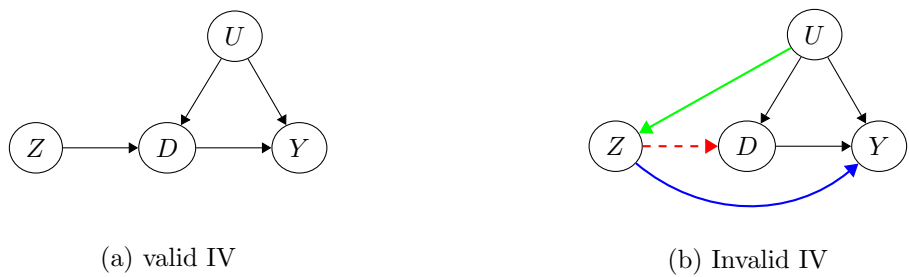

However, all of the above three core IV assumptions might be violated in large-scale genetics data, as shown in Figure \ref{fig:dag}(b). Among the three core IV assumptions \ref{ass: IV relevance}-\ref{ass: exclusion restriction}, only IV relevance assumption \ref{ass: IV relevance} is empirically testable by selecting genetic variants associated with the exposure, while IV independence assumption \ref{ass: IV independence} and exclusion restriction assumption \ref{ass: exclusion restriction} cannot be empirically verified in general \citep{davey2003mendelian, lawlor2008mendelian, sanderson2022mendelian}. The  near violation of the testable assumption \ref{ass: IV relevance} may happen when genetic variants exhibit weak associations with the exposure, leading to the potential weak IV bias \citep{staiger1994instrumental, stock2002survey, burgess2011avoiding, andrews2019weak}. The violation of assumption \ref{ass: IV independence} may arise due to the presence of population stratification, assortative mating and dynastic effect \citep{lawlor2008mendelian, brumpton2020avoiding, sanderson2021use}. The violation of assumption \ref{ass: exclusion restriction} may occur due to the widespread horizontal pleiotropy, where the genetic variant influences the outcome through other biological pathways that do not involve the exposure in view \citep{lawlor2008mendelian,sivakumaran2011abundant,solovieff2013pleiotropy, hemani2018evaluating}. Recently, a number of MR methods have been proposed for the identification, estimation and inference of the causal effect of interest when one or more of the three core IV assumptions are potentially violated \citep{bowden2015mendelian,bowden2017framework,verbanck2018detection, guo2018confidence,zhao2020statistical,sun2023semiparametric,liu2023mendelian,yao2024deciphering, zhang2025mr}.  For a more comprehensive review of identification and inference with invalid IVs, see \citet{kang2024identification}.

\section{Causal Estimands in the Potential Outcomes Framework}
\label{sec: estimand}

\subsection{Definition of individual treatment effect and average treatment effect}

Under the potential outcomes framework \citep{neyman1923applications,rubin1974estimating,rubin2005causal}, let $D_i\in\{0,1\}$ denote the binary treatment status (also referred to as the exposure) for subject $i$, and $Y_i(d)$ denote the potential outcome for subject $i$ if we set $D_i=d\in\{0,1\}$. The \textit{individual treatment effect} (ITE) \citep{neyman1923applications, rubin1974estimating} for subject $i$ is defined as
\begin{equation*}
    \text{ITE}_{i} = Y_i(1) - Y_i(0), \label{eq: ITE}
\end{equation*}
which quantifies the difference in the outcome for subject $i$ under treatment versus no treatment. The ITE is generally not identifiable since we cannot observe both $Y_i(1)$ and $Y_i(0)$ for the same subject $i$ at one time \citep{holland1986statistics,miguel2023causal}, a  concept known as the fundamental problem of causal inference. 

The \textit{average treatment effect} (ATE) \citep{rubin1974estimating, angrist1995identification} is defined as 
\begin{equation*}
    \text{ATE} = \mathbb{E}\left[ Y_i(1)-Y_i(0) \right], \label{eq: ATE}
\end{equation*}
which measures the difference in mean outcomes had everyone been treated versus had everyone been untreated. Likewise, for a continuous treatment, we can also define similar treatment effect for any two distinct levels: $d,d'$. Let $Y_i$ be the observed outcome for subject $i$. Under the following assumptions: (1) the consistency assumption, i.e., $Y_i=Y_i(d)$ for $d\in\{0,1\}$, and (2) the random assignment of treatment status, i.e., $\{Y_i(0),Y_i(1)\}\independent D_i$, the ATE can be identified as follows \citep{angrist1995identification,hirano2003efficient}:
\begin{equation*}
    \text{ATE} = \mathbb{E}\left[ Y_i(1)\right]- \mathbb{E}\left[Y_i(0) \right] = \mathbb{E}\left[ Y_i(1)|D_i=1\right]- \mathbb{E}\left[Y_i(0)|D_i=0 \right] = \mathbb{E}\left[ Y_i|D_i=1\right]- \mathbb{E}\left[Y_i|D_i=0 \right],
\end{equation*}
where the second equation holds because of the random assignment of treatment status, and the third equation holds because of the consistency assumption. However, when the treatment status is not randomized, the ATE cannot be identified using the above formula, because $ \mathbb{E}\left[ Y_i(d)\right] \neq \mathbb{E}\left[ Y_i(d)|D_i=d'\right]$ for $d,d'\in\{0,1\}$. 

\subsection{Constant treatment effect}

Consider a binary instrument variable (IV) $Z_i\in\{0,1\}$. Let $D_i(z)$ denote the binary treatment status for subject $i$ when the IV is set to $Z_i=z\in\{0,1\}$, and $Y_i(z,d)$ denote the potential outcome for subject $i$ if we set $Z_i=z\in\{0,1\}$ and $D_i=d\in\{0,1\}$. Then, the core IV assumptions \ref{ass: IV relevance}-\ref{ass: exclusion restriction} can be stated as:
\begin{assumptionp}{A1'}[IV relevance]
    $D_i \not\!\perp\!\!\!\!\perp Z_i$. \label{ass: IV relevance '}
\end{assumptionp}
\begin{assumptionp}{A2'}[IV independence]
    $\{Y_i(0,D_i(0)), Y_i(1,D_i(1)),D_i(0),D_i(1)\}\independent Z_i$. \label{ass: IV independence '}
\end{assumptionp}
\begin{assumptionp}{A3'}[Exclusion restriction]
    $Y_i(0,d)=Y_i(1,d)=Y_i(d)$ for $d\in\{0,1\}$. \label{ass: exclusion restriction '}
\end{assumptionp}

However, the above assumptions \ref{ass: IV relevance '}-\ref{ass: exclusion restriction '} alone are insufficient for the point identification of the causal effect, and hence a fourth assumption is required \citep{miguel2023causal}. One such assumption is the following constant treatment effect (CTE) assumption \citep{haavelmo1944probability, christ1966econometric, goldberger1972structural, hernan2006instruments, miguel2023causal}:
\begin{assumptionp}{A4.1}[Constant treatment effect]
    $Y_i(1)-Y_i(0)=\beta$ for all subjects $i$. \label{ass: Homogeneous effect}
\end{assumptionp}

Let $Y_i(0)=y_0+\varepsilon_i$, where $y_0=\mathbb{E}[Y_i(0)]$, then the observed outcome $Y_i$ can be written as the following model \citep{haavelmo1944probability, goldberger1972structural, wooldridge2010econometric}:
\begin{equation*}
    Y_i = y_0 + \beta D_i + \varepsilon_i.
\end{equation*}
Since $D_i$ might be correlated with $\varepsilon_i$, regressing the outcome $Y_i$ on the treatment $D_i$ does not consistently estimate $\beta$. However, under the IV independence assumption \ref{ass: IV independence '}, $\varepsilon_i$ should be independent of the instrument $Z_i$, implying $\mathbb{E}[\varepsilon_i|Z_i=0]=\mathbb{E}[\varepsilon_i|Z_i=1]$. Substituting $\varepsilon_i=Y_i-y_0-\beta D_i$ and solving for  $\beta$, it can be shown that the CTE $\beta$ equals the following usual IV estimand $\beta_{\text{IV}}$ \citep{wald1940fitting, miguel2023causal}:
\begin{equation}
    \beta_{\text{IV}} = \frac{\mathbb{E}[Y_i|Z_i=1] - \mathbb{E}[Y_i|Z_i=0]}{\mathbb{E}[D_i|Z_i=1]-\mathbb{E}[D_i|Z_i=0]}. \label{eq: usual IV estimand}
\end{equation}

\begin{figure}[H]
    \centering
    \begin{tikzpicture}[scale=2]
\tikzset{-}
\draw[->] (0,0) -- (2.7,0) node[below] {\( D_i \)};
\draw[->] (0,0) -- (0,2.5) node[left] {\( Y_i \)};

\coordinate (A) at (0.5,0.5);
\coordinate (B) at (2,2);

\draw[dotted, line width =1pt] (2,0.5) -- (0,0.5) node[left] {\( \mathbb{E}[Y_i | Z_i = 0] \)};
\draw[dotted, line width =1pt] (A) -- (0.5,0) node[below] {\( \mathbb{E}[D_i | Z_i = 0] \)};
\draw[dotted, line width =1pt] (B) -- (0,2) node[left] {\( \mathbb{E}[Y_i | Z_i = 1] \)};
\draw[dotted, line width =1pt] (B) -- (2,0) node[below] {\( \mathbb{E}[D_i | Z_i = 1] \)};

\draw[black,  thick, line width =2pt] (A) -- (B);

\fill[black] (A) circle (1.5pt);
\fill[black] (B) circle (1.5pt);

\draw [blue, ->, line width =2pt] (1.0,0.5) arc (0:45:0.5);
    \node at (1.2,0.7) {\color{blue}$\beta_{\text{IV}}$};

\end{tikzpicture}
    \caption{Graphical illustration of the usual IV estimand $\beta_{\text{IV}}$, represented by the slope of the solid line.}
    \label{fig: IV estimand}
\end{figure}
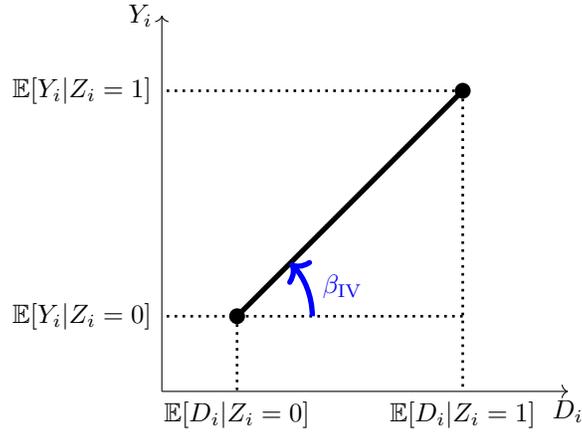

As illustrated in Figure \ref{fig: IV estimand}, the usual IV estimand $\beta_{\text{IV}}$ is the slope of the line that captures the relationship between the expected outcome $\mathbb{E}[Y_i|Z_i]$ and the expected treatment $\mathbb{E}[D_i|Z_i]$ conditional on two levels of the IV $Z_i=z\in\{0,1\}$.

\begin{conclusion}
    Under assumptions \ref{ass: IV relevance '}-\ref{ass: exclusion restriction '} and \ref{ass: Homogeneous effect}, the usual IV estimand \eqref{eq: usual IV estimand} identifies the constant treatment effect.
\end{conclusion}

\subsection{Average treatment effect on the treated}

In this section, we consider the following additive homogeneity assumption \citep{robins1994correcting,hernan2006instruments, miguel2023causal}, which is weaker than assumption \ref{ass: Homogeneous effect} and only requires that the average treatment effect is the same across different levels of $Z_i$ for both treated and untreated groups, i.e.,
\begin{assumptionp}{A4.2}[Additive homogeneity]
    $\mathbb{E}[Y_i(1)-Y_i(0)|D_i=d,Z_i=1]=\mathbb{E}[Y_i(1)-Y_i(0)|D_i=d,Z_i=0]$ for $d\in\{0,1\}$. \label{ass: additive homogeneity}
\end{assumptionp}
For binary treatment $D_i$ and binary IV $Z_i$, we can express the average treatment effect among the treated across different levels of $Z_i$ using the following saturated additive structural mean model \citep{robins1994correcting, hernan2006instruments, miguel2023causal}:
\begin{equation*}
    \mathbb{E}[Y_i(1)-Y_i(0)|D_i=1,Z_i] =  \beta_0 + \beta_1 Z_i.
\end{equation*}
With the consistency assumption, the above model can be re-written as $\mathbb{E}[Y_i-Y_i(0)|D_i,Z_i]=D_i(\beta_0+\beta_1Z_i)$ \citep{miguel2023causal}.
Here, $\beta_0$ represents the average treatment effect among the treated individuals with $Z_i=0$, and $\beta_0+\beta_1$ represents the average treatment effect among the treated individuals with $Z_i=1$. The additive homogeneity assumption \ref{ass: additive homogeneity} implies $\beta_1=0$, and then the parameter $\beta_0$ corresponds to the \textit{average treatment effect on the treated} (ATT) \citep{heckman1985alternative, robins1994correcting, hernan2006instruments}: 
\begin{equation*}
    \text{ATT} = \mathbb{E}\left[ Y_i(1)-Y_i(0) | D_i = 1 \right].\label{eq: ATT}
\end{equation*}
Under assumption \ref{ass: additive homogeneity},  $\mathbb{E}[Y_i(0)|Z_i=z]= \mathbb{E}[Y_i-D_i\beta_0 |Z_i=z]$. Under the IV independence assumption \ref{ass: IV independence '},  $\mathbb{E}[Y_i(0)|Z_i=0]= \mathbb{E}[Y_i(0)|Z_i=1]$. By solving the equation $\mathbb{E}[Y_i-D_i\beta_0 |Z_i=0]=\mathbb{E}[Y_i-D_i\beta_0 |Z_i=1]$, the parameter $\beta_0$ equals the usual IV estimand $\beta_{\text{IV}}$ defined in equation \eqref{eq: usual IV estimand}. More recently, \citet{liu2023mendelian} and \citet{liu2025quasi} have further investigated the identification of the ATT under potential violations of core IV assumptions.

\begin{conclusion}
    Under assumptions \ref{ass: IV relevance '}-\ref{ass: exclusion restriction '} and \ref{ass: additive homogeneity}, the usual IV estimand \eqref{eq: usual IV estimand} identifies the average treatment effect on the treated.
\end{conclusion}

\subsection{Local average treatment effect}

An alternative fourth identification assumption is the following monotonicity assumption \citep{angrist1995identification, angrist1996identification}:
\begin{assumptionp}{A4.3}[Monotonicity]
    $D_i(1)\geq D_i(0)$ for all subjects $i$.  \label{ass: monotonicity}
\end{assumptionp}
Then, under assumptions \ref{ass: IV relevance '}-\ref{ass: exclusion restriction '} and \ref{ass: monotonicity}, the usual IV estimand $\beta_{\text{IV}}$ in equation \eqref{eq: usual IV estimand} identifies the \textit{local average treatment effect} (LATE) in the subgroup of compliers (i.e., subjects with $D_i(0)=0$ and $D_i(1)=1$) \citep{angrist1995identification, angrist1996identification}, which is defined as:
\begin{equation*}
    \text{LATE} = \mathbb{E}\left[ Y_i(1)-Y_i(0) | \text{Compliers} \right]. \label{eq: LATE}
\end{equation*}

For binary IV $Z_i$ and binary treatment status $D_i$, the entire population is divided into four latent subgroups, known as compliance types \citep{angrist1995identification, angrist1996identification}, as shown in the following table:
\begin{table}[H]
\caption{Four compliance types based on the values of $D_i(z)$ for $z\in\{0,1\}$.}
    \centering
    \begin{tabular}{c|c|c}
    \hline
    & \( Z_{i}=0 \)  & \( Z_{i}=1 \)  \\
    \hline
    Complier & \( D_{i}(0)=0 \) & \( D_{i}(1)=1 \) \\
    Always-taker & \( D_{i}(0)=1 \) & \( D_{i}(1)=1 \) \\
    Never-taker & \( D_{i}(0)=0 \) & \( D_{i}(1)=0 \) \\
    Defier & \( D_{i}(0)=1 \) & \( D_{i}(1)=0 \) \\
    \hline
    \end{tabular}
    \label{tab: compliance}
\end{table}

The compliance type of subject $i$ is generally latent since we cannot observe both $D_i(0)$ and $D_i(1)$ at one time. Under the monotonicity assumption \ref{ass: monotonicity} there are no defiers in the population. Additionally, the IV $Z_i$ has no effect on $Y_i$ in the subgroups of always-takers or never-takers, as the treatment status $D_i$ is fixed across different levels of $Z_i$ in these two subgroups. Therefore, the IV $Z_i$ can only affect the outcome $Y_i$ in the subgroup of compliers, meaning that the usual IV estimand identifies the LATE in compliers.

\begin{conclusion}
    Under assumptions \ref{ass: IV relevance '}-\ref{ass: exclusion restriction '} and \ref{ass: monotonicity}, the usual IV estimand \eqref{eq: usual IV estimand} identifies the local average treatment effect in compliers.
\end{conclusion}

\subsection{Identification of average treatment effect}

Let $U_i$ denote the unmeasured confounders. \citet{wang2018bounded} proposes the following two no-interaction assumptions for the identification of ATE:
\begin{assumptionp}{A4.4}[No additive $U_i-Z_i$ interaction]
    There is no additive $U_i-Z_i$ interaction in $\mathbb{E}[D_i|Z_i,U_i]$, that is, $\mathbb{E}[D_i|Z_i=1,U_i]-\mathbb{E}[D_i|Z_i=0,U_i]=\mathbb{E}[D_i|Z_i=1]-\mathbb{E}[D_i|Z_i=0]$. \label{ass: no U-Z interact}
\end{assumptionp}

\begin{assumptionp}{A4.5}[No additive $U_i-d$ interaction]
    There is no additive $U_i-d$ interaction in $\mathbb{E}[Y_i(d)|U_i]$, that is, $\mathbb{E}[Y_i(1)-Y_i(0)|U_i]=\mathbb{E}[Y_i(1)-Y_i(0)]$. \label{ass: no U-d interact}
\end{assumptionp}
Intuitively, assumption \ref{ass: no U-Z interact} rules out  modification  of instrument-treatment association by $U_i$ on the additive scale, whereas assumption \ref{ass: no U-d interact} rules out modification of the effect of the treatment on the outcome by $U_i$ on the additive scale. In addition, \citet{wang2018bounded} also imposes the following assumption for confounding control:
\begin{assumptionp}{A5}[Sufficiency of $U_i$ for confounding control]
    $Y_i(d)\independent (D_i,Z_i) ~|~ U_i$. \label{ass: confounding control}
\end{assumptionp}
Assumption \ref{ass: confounding control} requires that, conditional on the unmeasured confounders $U_i$, the potential outcomes are independent of the treatment status and the instrument. This assumption is originally formulated by \citet{richardson2014ace}.

\begin{conclusion}
    Under assumptions \ref{ass: IV relevance '}-\ref{ass: exclusion restriction '} and \ref{ass: confounding control}, together with either assumption \ref{ass: no U-Z interact} or \ref{ass: no U-d interact},  the usual IV estimand \eqref{eq: usual IV estimand} identifies the average treatment effect.
\end{conclusion}

\subsection{Comparisons of the causal estimands}

We compare the four causal estimands (CTE, ATT, LATE and ATE) in Table \ref{tab: comparison of estimands}. The identification of all four causal estimands requires assumptions \ref{ass: IV relevance '}-\ref{ass: exclusion restriction '}, which are therefore referred to as the \textit{core IV assumptions} \citep{angrist1996identification, baiocchi2014instrumental}. However, these core IV assumptions \ref{ass: IV relevance '}-\ref{ass: exclusion restriction '} are insufficient for point identification, and the four causal estimands differ in their additional identification assumption \citep{miguel2023causal}. The CTE relies on the strong constant treatment effect assumption \ref{ass: Homogeneous effect}, which posits that the treatment effect is the same across all subjects. By contrast, the ATT relies on the additive homogeneity assumption \ref{ass: additive homogeneity}, a weaker  identification assumption than \ref{ass: Homogeneous effect},  and corresponds to the average treatment effect among those who actually received the treatment. The LATE is identified by imposing the monotonicity assumption \ref{ass: monotonicity} as the additional identification assumption, and is interpreted as the average treatment effect in the subgroup of  compliers, i.e., subjects who would receive the treatment if assigned to it and not receive it otherwise. Finally, the ATE, which captures the average treatment effect for the entire population, is identified under either the no-interaction assumption \ref{ass: no U-Z interact} or \ref{ass: no U-d interact}, together with the confounding control assumption \ref{ass: confounding control}.

\begin{table}[H]
\caption{Comparison of identification assumptions and interpretations for CTE, ATE, ATT, and LATE.}
\begin{tabular}{cll}
\hline
Causal   estimand & Identification assumptions                                                                                                                        & Causal interpretation                                                                                                                                                                       \\ \hline
CTE               & \begin{tabular}[c]{@{}l@{}}IV relevance \ref{ass: IV relevance '}; \\ IV independence \ref{ass: IV independence '};\\ Exclusion restriction \ref{ass: exclusion restriction '};\\ Constant treatment effect \ref{ass: Homogeneous effect}. \end{tabular} & \begin{tabular}[c]{@{}l@{}}Constant treatment effect of   \\ the treatment versus control \\ across all subjects.\end{tabular}                                                          \\ \hline
ATT               &\begin{tabular}[c]{@{}l@{}}IV relevance \ref{ass: IV relevance '}; \\ IV independence \ref{ass: IV independence '};\\ Exclusion restriction \ref{ass: exclusion restriction '};\\ Additive homogeneity \ref{ass: additive homogeneity}. \end{tabular}            & \begin{tabular}[c]{@{}l@{}}Average treatment effect of   \\ the treatment versus control \\ specifically for subjects that \\ actually received the treatment.\end{tabular}               \\ \hline
LATE              & \begin{tabular}[c]{@{}l@{}}IV relevance \ref{ass: IV relevance '}; \\ IV independence \ref{ass: IV independence '};\\ Exclusion restriction \ref{ass: exclusion restriction '};\\ Monotonicity \ref{ass: monotonicity}. \end{tabular}                    & \begin{tabular}[c]{@{}l@{}}Average treatment effect of   \\ the treatment versus control \\ specifically for the compliers. 
\end{tabular} \\ \hline
ATE               & \begin{tabular}[c]{@{}l@{}}IV relevance \ref{ass: IV relevance '}; \\ IV independence \ref{ass: IV independence '};\\ Exclusion restriction \ref{ass: exclusion restriction '};\\ No-interaction \ref{ass: no U-Z interact} or \ref{ass: no U-d interact};\\ Confounding control \ref{ass: confounding control}. \end{tabular} & \begin{tabular}[c]{@{}l@{}}Average treatment effect of   \\ the treatment versus control \\ for the entire population.\end{tabular}                                                          \\ \hline
\end{tabular}
\label{tab: comparison of estimands}
\end{table}

\section{Causal Estimand Defined in the ALICE Model}
\label{sec: ALICE}
\subsection{Definition of the ALICE model }

Having introduced the key causal estimands (CTE, ATT, LATE, and ATE) and their identification under IV assumptions, we now turn to a specific causal model that has become the workhorse of Mendelian randomization studies. The \underline{A}dditive \underline{LI}near \underline{C}onstant \underline{E}ffects (ALICE) model \citep{holland1988causal} formalizes the constant treatment effect assumption \ref{ass: Homogeneous effect} introduced in Section \ref{sec: estimand}, providing a simple yet widely used framework for characterizing causal effects in MR. Let $D_i\in\mathbb{R}$ denote the exposure of subject $i$, and $\bm{Z}_{i}=(Z_{i1},\cdots,Z_{ip})^\top\in\mathbb{R}^p$ denote the vector of $p$ genetic instruments of subject $i$. Let $Y_i(\bm{z},d)\in\mathbb{R}$ denote the continuous potential outcome if subject $i$ had $\bm{Z}_i=\bm{z}=(z_1,\cdots,z_p)^\top$ and $D_i=d$. Then, for two possible values of instruments $\bm{z},\bm{z}'$ and the exposure $d,d'$, we assume the following model \citep{holland1986statistics, kang2016instrumental, guo2018confidence}:
\begin{align}
\begin{split}
    Y_i(\bm{z}',d') - Y_i(\bm{z},d) &= \beta (d'-d)  + \sum_{j=1}^p \psi_j (z_j'-z_j) , \\
    \mathbb{E}[Y_i(\bm{0},0)|\bm{Z}_i] & = \sum_{j=1}^p  \phi_j Z_{ij} ,  \label{eq: ALICE for individual}
\end{split}
\end{align}
where $\beta\in\mathbb{R}$ is the primary causal parameter of interest, representing the constant effect of a one-unit change in the exposure on the outcome across all subjects in the whole population. The parameter $\psi_j\in\mathbb{R}$  quantifies the degree of violation of the exclusion restriction assumption \ref{ass: exclusion restriction} for $j$th instrument, capturing the direct effects of the instrument on the potential outcome. The parameter $\phi_j\in \mathbb{R}$ quantifies the degree of violation of the IV independence assumption \ref{ass: IV independence} for $j$th genetic instrument. Under the IV independence assumption \ref{ass: IV independence}, the instruments $\bm{Z}_i$ should be independent of the baseline potential outcome $Y_i(\bm{0}, 0)$ in the absence of confounding. However, in model \eqref{eq: ALICE for individual}, the relationship between $\bm{Z}_i$ and $Y_i(\bm{0}, 0)$ is modeled through $\phi_1,\cdots,\phi_p$, allowing for potential violations of assumption \ref{ass: IV independence} \citep{angrist1996identification, kang2016instrumental, guo2018confidence}. Let $\pi_j=\psi_j+\phi_j$ for $j=1,\cdots,p$, and $\varepsilon_i=Y_i(0,\bm{0})-\mathbb{E}[Y_i(0,\bm{0})|\bm{Z}_i]$, then under the consistency assumption, we have the following observed outcome model \citep{small2007sensitivity,kang2016instrumental,guo2018confidence}:
\begin{equation}
    Y_i = \beta D_i + \sum_{j=1}^p \pi_j Z_{ij} + \varepsilon_i, \quad \quad   \mathbb{E}(\varepsilon_i|\bm{Z}_i) = 0.\label{eq: outcome model}
\end{equation}
The causal effect $\beta$ in model \eqref{eq: outcome model} cannot be estimated by directly fitting a usual linear regression because the exposure $D_i$ might be correlated with the error term $\varepsilon_i$. Moreover, in model \eqref{eq: outcome model},  the parameter $\pi_j \in \mathbb{R}$ encodes the degrees of violation of assumptions \ref{ass: IV independence} and \ref{ass: exclusion restriction} for $j$th genetic instrument. Specifically, if the $j$th genetic instrument satisfies both the exclusion restriction assumption and IV independence assumption, then $\pi_j=0$; otherwise, if $\pi_j\neq0$, the $j$th genetic instrument violates at least one of the exclusion restriction assumption or IV independence assumption \citep{kang2016instrumental, guo2018confidence, windmeijer2021confidence, guo2023causal, sun2023semiparametric, kang2024identification, zhang2025mr}. Therefore, we say the $j$th genetic instrument is a valid IV if $\pi_j=0$, and an invalid IV if $\pi_j\neq 0$. In Section \ref{sec: invalid IV}, we will discuss the identification and inference in the presence of invalid IVs under the ALICE model framework.

\begin{remark}    \citet{kang2016instrumental} extends model \eqref{eq: ALICE for individual} to incorporate heterogeneous causal effect as follows: 
\begin{equation*}
    Y_i(\bm{z}',d') - Y_i(\bm{z},d) =  \beta_i (d'-d) + \sum_{j=1}^p \psi_j (z_j'-z_j),
\end{equation*}
where $\beta_i$ is the individual causal effect of subject $i$. Let $\beta = \mathbb{E}[\beta_i]$ be the average causal effect, the observed outcome model \eqref{eq: outcome model} becomes
\begin{equation*}
    Y_i = \beta D_i + \sum_{j=1}^p \pi_j Z_{ij} + (\beta_i-\beta) D_i + \varepsilon_i, \quad \quad   \mathbb{E}(\varepsilon_i|\bm{Z}_i) = 0.
\end{equation*}
This model reduces to the constant causal effect model \eqref{eq: outcome model} if $(\beta_i - \beta)$ is independent of $D_i$ given $\mathbf{Z}_i$ \citep{kang2016instrumental}.
\end{remark}

\subsection{ALICE model is widely used in MR studies}

To model the relationship between a continuous exposure and genetic instruments, we further consider a linear model between the exposure $D_i$ and the genetic instruments $\bm{Z}_i$ \citep{angrist1996identification, small2007sensitivity, guo2018confidence}:
\begin{equation}
    D_i = \sum_{j=1}^p \gamma_j Z_{ij} +  \delta_i, \quad \quad  \mathbb{E}(\delta_i|\bm{Z}) = 0, \label{eq: exposure model}
\end{equation}
where $\gamma_j$ represents the IV strength of $j$th genetic instrument.

Note that the error term $\delta_i$ in the exposure model \eqref{eq: exposure model} might be correlated with the error term $\varepsilon_i$ in the outcome model \eqref{eq: outcome model} due to unmeasured confounders. By plugging in the exposure model \eqref{eq: exposure model} into the outcome model \eqref{eq: outcome model}, we can obtain the reduced-form model for the outcome \citep{small2007sensitivity,guo2018confidence}:
\begin{equation}
    Y_i = \sum_{j=1}^p \Gamma_j Z_{ij} + e_i, \quad\quad \mathbb{E}(e_i|\bm{Z}_i) = 0, \label{eq: reduced-form model}
\end{equation}
where $\Gamma_j=\beta\gamma_j+\pi_j$, and $e_i=\beta \delta_i + \varepsilon_i$.

Most summary-level MR methods for continuous outcomes build upon the ALICE model. For a single genetic instrument $j$, according to the equation  $\Gamma_j=\beta\gamma_j+\pi_j$, the ratio estimand is defined as follows \citep{burgess2013mendelian, slob2020comparison}:
\begin{equation}
    \beta_j = \frac{\Gamma_j}{\gamma_j} = \beta+ \frac{\pi_j}{\gamma_j}. \label{eq: ratio estimand}
\end{equation}
When $j$th genetic instrument is a valid IV (i.e., $\pi_j = 0$), the ratio estimand $\beta_j$ equals the causal effect $\beta$ in the ALICE model. In summary-level MR analysis, the ratio estimate of $j$th SNP is defined as $\widehat{\beta}_j=\widehat{\Gamma}_j/\widehat{\gamma}_j$, where $\widehat{\Gamma}_j$ and $\widehat{\gamma}_j$ are marginal estimates of $\Gamma_j$ and $\gamma_j$ in genome-wide association studies (GWAS) summary statistics.

\subsection{Practical limitations and interpretational caveats of the ALICE model}

Although the ALICE model provides a useful framework for causal inference with potentially invalid instrumental variables, it is subject to several important limitations and interpretational caveats. First, the causal effect defined in the ALICE model should be interpreted as a constant treatment effect  of a one-unit increase in the exposure on the outcome \citep{holland1986statistics, small2007sensitivity, kang2016instrumental}. However, the constant treatment effect assumption may not hold in real-world settings where treatment effects may vary across subjects \citep{angrist2004treatment, powers2018some, kunzel2019metalearners}. Second, the ALICE model assumes linearity not only in the causal effect but also in the violation of core IV assumptions. This linearity assumption might be violated when the underlying relationships are nonlinear, such as in complex genetic architectures \citep{veitia2013gene, guindo2021impact, sun2023semiparametric}. Therefore, when applying the ALICE model, it is essential to carefully assess the plausibility of its assumptions within the context of the study and to interpret the resulting estimates with appropriate caution.

\section{Identification and Inference in the Presence of Invalid IVs}
\label{sec: invalid IV}

\subsection{Additional identification assumption under the ALICE model}

When $j$th genetic IV is a valid instrument (i.e., $\pi_j=0$), the ALICE model in Section \ref{sec: ALICE} enables causal identification through the ratio $\Gamma_j/\gamma_j$. However, when invalid instruments are present without prior knowledge of IV validity status, the causal effect $\beta$ in model \eqref{eq: outcome model} becomes non-identifiable. This is because the parameters in models \eqref{eq: exposure model} and \eqref{eq: reduced-form model} should satisfy the following equation system:
\begin{equation}
    \Gamma_j=\beta\gamma_j+\pi_j,~~~~j=1,\cdots,p, \label{eq: equation system}
\end{equation}
where the IV-exposure associations $\bm{\gamma}=(\gamma_1,\cdots,\gamma_p)^\top$ and IV-outcome associations $\bm{\Gamma}=(\Gamma_1,\cdots,\Gamma_p)^\top$ can be identified using population ordinary least squares (OLS) through $\bm{\gamma}=\mathbb{E}(\bm{Z}_i \bm{Z}_i^\top)^{-1}\mathbb{E}(\bm{Z}_i D_i)$ and $\bm{\Gamma}=\mathbb{E}(\bm{Z}_i \bm{Z}_i^\top)^{-1}\mathbb{E}(\bm{Z}_i Y_i)$. Given $\bm{\gamma}$ and $\bm{\Gamma}$, there are $p$ equations with $p+1$ unknown parameters $\{\beta,\pi_1,\cdots,\pi_p\}$), resulting in an underdetermined equation system that precludes unique identification of $\{\beta, \pi_1,\cdots,\pi_p\}$, and renders models \eqref{eq: outcome model} and \eqref{eq: exposure model} under-identified.  Consequently, additional assumptions regarding $\bm{\pi}=(\pi_1,\cdots,\pi_p)^\top$ are required to address the identifiability issue. Below, we list three commonly adopted additional identification assumptions in the ALICE model.

\begin{assumptionp}{A6}[Instrument strength independent of the direct effect (InSIDE)]
    The IV-exposure association $\gamma_j$ is asymptotically independent of the degree of IV invalidity $\pi_j$ as the number of genetic IVs $p$ goes to infinity. \label{ass: InSIDE}
\end{assumptionp}

From the equation system \eqref{eq: equation system}, we have 
    \begin{equation*}
        \frac{\text{Cov}(\bm{\Gamma},\bm{\gamma})}{\text{Var}(\bm{\gamma})} = \beta + \frac{\text{Cov}(\bm{\pi},\bm{\gamma})}{\text{Var}(\bm{\gamma})}.
    \end{equation*}
Under the InSIDE assumption, $\text{Cov}(\bm{\pi},\bm{\gamma})\to0$ as $p\to \infty$, yielding the identification of $\beta$ \citep{kolesar2015identification, bowden2015mendelian}. The InSIDE assumption has been  adopted in some summary-level MR methods,  for example, MR-Egger \citep{bowden2015mendelian}, random-effects inverse-variance weighed (IVW) method \citep{bowden2017framework}, and MR using the Robust Adjusted Profile Score (MR-RAPS) \citep{zhao2020statistical}.

\begin{assumptionp}{A7}[Majority rule]
    The number of valid genetic IVs is more than half of the relevant genetic IVs. \label{ass: majority}
\end{assumptionp}

The majority rule assumption is a sufficient condition for the identification of $\beta$ under the ALICE model framework \citep{han2008detecting, kang2016instrumental}. Formally, let $\mathcal{S}=\{j: \gamma_j\neq 0\}$ denote the set of all relevant genetic IVs with non-zero IV-exposure associations, and $\mathcal{V}=\{j\in\mathcal{S}:\gamma_j\neq0 ~~\text{and}~~\pi_j=0\}$ denote the set of all valid genetic IVs, then the majority rule assumption can be  expressed as $|\mathcal{V}|>\frac{1}{2}|\mathcal{S}|$. Under the majority rule assumption, more than half of the ratio estimand $\beta_j$ defined in equation \eqref{eq: ratio estimand} equal the true causal effect $\beta$, since they arise from valid instruments with $\pi_j = 0$ \citep{bowden2016consistent}. A natural identification strategy is therefore to find the median of $\{\beta_j\}_{j \in \mathcal{S}}$ \citep{bowden2016consistent}. Some MR methods based on the majority rule assumption include Some Invalid Some Valid IV Estimator (sisVIVE) \citep{kang2016instrumental}, weighted median method \citep{bowden2016consistent}, and MR Pleiotropy RESidual Sum and Outlier test (MR-PRESSO) \citep{verbanck2018detection}. 

\begin{assumptionp}{A8}[Plurality rule]
    Valid genetic IVs form the largest group among relevant genetic IVs based on the ratio of IV-outcome association to IV-exposure association. \label{ass: plurality}
\end{assumptionp}

As shown in \citet{guo2018confidence}, the plurality rule assumption is weaker than the majority rule assumption, and is a sufficient condition for the identification of causal effect $\beta$ under the ALICE model framework. Formally, the plurality rule assumption can be expressed as $|\mathcal{V}|>\max_{c\neq 0}|\{j\in \mathcal{S}:\frac{\pi_j}{\gamma_j}=c\}|$. Under this assumption, the true causal effect $\beta$ corresponds to the mode of the distribution of $\{\beta_j\}_{j \in \mathcal{S}}$ \citep{hartwig2017robust}. Thus, the identification of the causal effect $\beta$ can be achieved by detecting the largest group of ratio estimands, either by direct mode estimation \citep{hartwig2017robust} or via voting-based procedures \citep{guo2018confidence, yao2024deciphering}. The plurality rule assumption is also termed as the ZEro Modal Pleiotropy Assumption (ZEMPA) \citep{hartwig2017robust}, and is adopted in MR methods including the mode-based estimation \citep{hartwig2017robust}, Two-Stage Hard Thresholding (TSHT) \citep{guo2018confidence}, MRMix \citep{qi2019mendelian}, the contamination mixture method \citep{burgess2020robust}, Confidence Interval method for Instrumental Variable (CIIV) \citep{windmeijer2021confidence}, and MR with valid IV Selection and Post-selection Inference (MR-SPI) \citep{yao2024deciphering}.

\begin{remark}
Equation \eqref{eq: equation system} demonstrates that, in the presence of unknown instrument invalidity, the causal effect $\beta$ in the ALICE model is generally not identifiable. Assumptions \ref{ass: InSIDE}-\ref{ass: plurality} restore identification by imposing different constraints on $\bm{\pi}$, which encodes the degree of violation of assumptions \ref{ass: IV independence} and \ref{ass: exclusion restriction}. Because assumptions \ref{ass: InSIDE}-\ref{ass: plurality} cannot be empirically tested with data, a common practice is to employ multiple MR methods relying on different assumptions as sensitivity analyses to evaluate the robustness of MR findings.
\end{remark}

\subsection{Alternative identification strategies beyond the ALICE model framework}

\citet{sun2023semiparametric} considers the following model under the potential outcomes framework:
\begin{equation*}
    Y_i(\bm{z},d') - Y_i(\bm{0},d) = \beta (d'-d) + \psi(\bm{z}),
\end{equation*}
where $\psi(\cdot)$ is an unknown function that satisfies $\psi(\bm{0})=0$, which allows for arbitrary interactions among the direct effects of the instruments on the outcome. The ALICE model is a special case of the above model by specifying $\psi(\bm{z})=\sum_{j=1}^p \psi_j z_j$ and $\mathbb{E}[Y_i(\bm{0},0)|\bm{Z}_i]  = \sum_{j=1}^p  \phi_j Z_{ij}$ \citep{sun2023semiparametric}. Under this model, the set of valid instruments is defined as the index set $\mathcal{V}\subseteq \{1,\cdots,p\}$ such that $\psi(\bm{\bm{Z}_i})=\psi(\bm{Z}_{i,-\mathcal{V}})$ and $\mathbb{E}[Y_i(\bm{0},0)|\bm{Z}_i]  = \mathbb{E}[Y_i(\bm{0},0)|\bm{Z}_{i,-\mathcal{V}}]$ holds almost surely, where $\bm{Z}_{i,-\mathcal{V}}=(Z_{ij}:j\notin \mathcal{V})$ \citep{sun2023semiparametric}. When all $p$ instruments are mutually independent and there are at least $v$ valid instruments, \citet{sun2023semiparametric} shows that the causal effect $\beta$ in the above model is the unique solution to the following equation:
\begin{equation*}
    \mathbb{E}[\bm{h}^{[v]}(\bm{Z}_i)(Y_i-\beta D_i)]=\bm{0}, 
\end{equation*}
where the function $\bm{h}^{[v]}(\bm{Z}_i)\in\mathbb{R}^m$ with $m=\sum_{j=0}^{v-1} \binom{p}{j}$ represents all demeaned interactions involving at least $p-v+1$ instruments. For example, when there are $p=2$ instruments and there is at least $v=1$ valid instrument, then there is only one demeaned interaction $(Z_{i1}-\mu_1)(Z_{i2}-\mu_2)$ involving at least 2 instruments, where $\mu_1$ and $\mu_2$ are the expectations of $Z_{i1}$ and $Z_{i2}$, respectively. When $(Z_{i1}-\mu_1)(Z_{i2}-\mu_2)$ is associated with the exposure $D_i$, then $\beta$ is the unique solution to
\begin{equation*}
    \mathbb{E}[(Z_{i1}-\mu_1)(Z_{i2}-\mu_2)(Y_i-\beta D_i)] = 0.
\end{equation*}

\begin{remark}
    As discussed in \citet{kang2024identification}, \citet{sun2023semiparametric} uses higher-order interactions to create ``new" instruments from the $p$ instruments, which can capture possible nonlinear effects of instruments on the exposure. In contrast, \citet{guo2022robustness} applies machine learning algorithms to explore nonlinear effects of instruments on the exposure.
\end{remark}

\citet{tchetgen2021genius} proposes the MR G-Estimation under No Interaction with Unmeasured Selection (MR-GENIUS) approach that leverages heteroscedasticity in the exposure to identify the causal effect. Specifically, \citet{tchetgen2021genius} considers the following model:
\begin{align*}
    \mathbb{E}[Y_i|D_i,\bm{Z}_i,U_i] &= \beta_y(U_i) D_i + \alpha_y(U_i,\bm{Z}_i) + \eta_y(U_i), \\ 
    \mathbb{E}[D_i|\bm{Z}_i,U_i] &= \alpha_d(U_i,\bm{Z}_i) + \eta_d(U_i),
\end{align*}
where $\beta_y(\cdot)$ is an unspecified function of the unmeasured confounder $U_i$, which affects both the exposure $D_i$ and the outcome $Y_i$, and is independent of the instruments $\bm{Z}_i$. The terms $\eta_y(\cdot)$ and $\eta_d(\cdot)$ are two unspecified functions of $U_i$, and $\alpha_y(\cdot)$ and $\alpha_d(\cdot)$ are two unspecified functions of $(U_i,\bm{Z}_i)$ satisfying $\alpha_y(U,\bm{0})=\alpha_d(U,\bm{0})=0$. When the exposure $D_i$ is heteroscedastic, i.e., $\text{Var}(D_i|\bm{Z}_i)$ varies with the instruments $\bm{Z}_i$, \citet{tchetgen2021genius} shows that the average causal effect $\beta=\mathbb{E}[\beta_y(U_i)]$ in the above model is the unique solution to the following equation:
\begin{equation*}
    \mathbb{E}[(\bm{Z}_i-\mathbb{E}(\bm{Z}_i))(D_i-\mathbb{E}(D_i|\bm{Z}_i))(Y_i-\beta D_i)] = \bm{0}.
\end{equation*}

\begin{remark}
    As discussed in \citet{tchetgen2021genius}, MR-GENIUS might not perform well when $\text{Var}(D_i|\bm{Z}_i)$ is only weakly dependent on the instruments $\bm{Z}_i$. \citet{ye2024genius} extends MR-GENIUS to allow for many weak invalid instruments.
\end{remark}

By leveraging heteroscedasticity in the outcome, \citet{liu2023mendelian} proposes the Mendelian Randomization Mixed-Scale Treatment Effect Robust Identification (MR-MiSTERI) approach for the average treatment effect on the treated (ATT). In this section, we focus on the case where both the treatment and the possibly invalid genetic instrument are binary; see \citet{liu2023mendelian} for extensions. Specifically, MR-MiSTERI relies on the following three identification assumptions:

\begin{assumptionp}{B1}[Homogeneous ATT]
    The ATT does not vary with the possibly invalid IV on the additive scale, i.e., $\mathbb{E}[Y_i(z,d=1)-Y_i(z,d=0)~|~D_i=1,Z_i=z]=\beta$. \label{ass: Homogeneous ATT}
\end{assumptionp}

\begin{assumptionp}{B2}[Homogeneous confounding bias on the odds ratio scale]
    $\text{OR}(Y_i(0)=y_0,D_i=d~|~Z_i=z)=\exp(\xi d y_0)$, where $\xi$ quantifies the magnitude of confounding bias. \label{ass: Homogeneous confounding bias}
\end{assumptionp}

\begin{assumptionp}{B3}[Outcome heteroscedasticity]
    Define $\varepsilon_i=Y_i-\mathbb{E}(Y_i~|~D_i,Z_i)$ and suppose that $\varepsilon_i~|~D_i,Z_i\sim N(0,\sigma^2(Z_i))$, then  $\sigma^2(Z_i)$ must vary with the genetic instrument $Z_i$.  \label{ass: Outcome heteroscedasticity}
\end{assumptionp}

\begin{remark}
Assumption \ref{ass: Outcome heteroscedasticity} can be empirically testable through genome-wide variance quantitative trait loci (vQTL) analyses \citep{pare2010use, wang2019genotype}. As shown in \citet{pare2010use},  gene–gene (GxG) and/or gene–environment (GxE) interactions can result in genotype-dependent changes in trait variance, providing direct evidence for heteroscedasticity in quantitative traits.
\end{remark}

Under assumptions \ref{ass: Homogeneous ATT}-\ref{ass: Outcome heteroscedasticity}, the confounding bias parameter $\xi$ and the causal effect $\beta$ are uniquely identified by
\begin{align*}
    \xi &= \frac{D_i(Z_i=1)-D_i(Z_i=0)}{\sigma^2(Z_i=1)-\sigma^2(Z_i=0)}, \\
    \beta &= D_i(Z_i) - \frac{D_i(Z_i=1)-D_i(Z_i=0)}{\sigma^2(Z_i=1)-\sigma^2(Z_i=0)} \sigma^2(Z_i),~~Z_i=0,1,
\end{align*}
and the estimates for $\xi$ and $\beta$ can be obtained by replacing the unknown quantities with the sample counterparts in observed data.

\subsection{Inference for the causal effect}

Following \citet{kang2024identification}, inference for the causal effect in MR analysis can be broadly classified into two methodological paradigms: \textit{pointwise inference} and \textit{uniformly valid inference}. Pointwise inference constructs the confidence interval for the causal effect either by: (1) calculating the standard error of the causal effect estimate directly from asymptotic distribution or resampling techniques (e.g., bootstrap) \citep{bowden2015mendelian, bowden2016consistent, bowden2017framework, hartwig2017robust, qi2019mendelian, zhao2020statistical, burgess2020robust}; or (2) selecting valid instruments from candidate genetic variants  (e.g., using voting procedure or outlier detection test) and subsequently constructing confidence intervals using the selected subset \citep{guo2018confidence, verbanck2018detection, windmeijer2021confidence, yao2024deciphering}. The latter approach relies on the correct selection of valid IVs; when IV selection error occurs in finite samples, it might lead to poor coverage performance \citep{guo2023causal}.

The second paradigm, \textit{uniformly valid inference}, constructs CIs that remain robust to finite-sample instrument selection errors \citep{kang2022two,guo2023causal, yao2024deciphering, kang2024identification}. Specifically, \citet{kang2022two} proposes taking the union of confidence intervals constructed from subsets of instruments passing the $J$ test \citep{hansen1982large}; however,  this procedure is computationally costly when the number of candidate genetic IVs is large \citep{kang2024identification}. Alternatively, \citet{guo2023causal} and \citet{yao2024deciphering} first construct ``pseudo CIs" through grid-search using resampled IV-exposure and IV-outcome associations, and then construct the final robust CI by taking the union of these pseudo CIs across resamples. Uniformly valid inference generally constructs wider CIs than pointwise inference, which is a trade-off for the guaranteed finite-sample coverage level \citep{kang2024identification}.

\section{Weak Identification in the ALICE Model Framework}
\label{sec: weak IV}
\subsection{The presence of weak identification bias}
In this section, we examine the bias introduced by weak instruments, i.e., instruments that are only weakly associated with the exposure, in the ALICE model framework. We begin by assuming that all instruments satisfy assumptions \ref{ass: IV independence} and \ref{ass: exclusion restriction}, i.e., $\pi_j=0$ for all $j\in\{1,\cdots,p\}$ in model \eqref{eq: outcome model}. In this case, two-stage least squares (2SLS) is commonly employed to estimate the causal effect $\beta$
 \citep{angrist2009mostly, wooldridge2016introductory}. Specifically, in the first stage, we fit an OLS regression of the exposure $\bm{D}$  on the genetic instruments $\bm{Z}$ to obtain the following fitted exposure values $\widehat{\bm{D}}$:
\begin{equation*}
    \widehat{\bm{D}} = \bm{Z}\left(\bm{Z}^\top \bm{Z}\right)^{-1} \bm{Z}^\top\bm{D}.
\end{equation*}
In the second stage, the outcome $\bm{Y}$ is regressed on these fitted exposures $\widehat{\bm{D}}$ to obtain the following 2SLS estimator of the causal effect:
\begin{equation}
    \widehat{\beta}_{\text{2SLS}} = \left(\widehat{\bm{D}}^T\widehat{\bm{D}}\right)^{-1}\widehat{\bm{D}}^\top\bm{Y}. \label{eq: 2SLS}
\end{equation}
Assume that the error terms satisfy $\varepsilon_i\sim N(0,\sigma_\varepsilon^2)$, $\delta_i\sim N(0,\sigma_\delta^2)$, and denote $\text{Cov}(\delta_i, \varepsilon_i)=\sigma_{\delta,\varepsilon}$, \citet{rothenberg1984approximating} provides the following analytic expression for the bias of 2SLS estimator
\begin{equation}
    \mu\left(\widehat{\beta}_{\text{2SLS}} - \beta \right) = \frac{\sigma_\varepsilon}{\sigma_\delta} \frac{\eta_\varepsilon + \xi_{\varepsilon,\delta}/\mu}{1+ 2\eta_\delta/\mu+ \xi_{\delta,\delta}/\mu^2}, \label{eq: bias of 2SLS}
\end{equation}
where $\mu^2 = \bm{\gamma}^\top \bm{Z}^\top \bm{Z}\bm{\gamma}/\sigma_\delta^2$ is the \textit{concentration parameter} that measures the instrument strength \citep{stock2002survey, stock2002testing}. Here, $\eta_\varepsilon=(\sigma_\varepsilon\sqrt{\bm{\gamma}^\top \bm{Z}^\top \bm{Z}\bm{\gamma}})^{-1}\bm{\gamma}^\top \bm{Z}^\top \bm{\varepsilon} $ and $\eta_\delta=(\sigma_\delta\sqrt{\bm{\gamma}^\top \bm{Z}^\top \bm{Z}\bm{\gamma}})^{-1}\bm{\gamma}^\top \bm{Z}^\top \bm{\delta}$ are two standard normal random variables with correlation $\sigma_{\delta,\varepsilon}/(\sigma_{\varepsilon}\sigma_{\delta})$,  $\xi_{\varepsilon,\delta}=\bm{\delta}^\top\bm{Z}\left(\sigma_{\varepsilon}\sigma_{\delta}\bm{Z}^\top \bm{Z}\right)^{-1} \bm{Z}^\top \bm{\varepsilon}$ and $\xi_{\delta,\delta}=\bm{\delta}^\top\bm{Z}\left(\sigma_{\delta}^2\bm{Z}^\top \bm{Z}\right)^{-1} \bm{Z}^\top \bm{\delta}$ are two quadratic forms of normal random variables that do not depend on the sample size $n$.
\begin{remark}
From equation \eqref{eq: bias of 2SLS}, as the concentration parameter $\mu^2$ goes to infinity, $\mu\left(\widehat{\beta}_{\text{2SLS}} - \beta \right)$ has an asymptotic distribution of $N(0, \sigma_\varepsilon^2/\sigma_\delta^2)$ \citep{rothenberg1984approximating}. Therefore, the concentration parameter $\mu^2$ can be thought of as an effective sample size \citep{stock2002survey, stock2002testing}. When instruments are strong, the concentration parameter $\mu^2$ increases proportionally to the sample size $n$ \citep{andrews2005inference}.
\end{remark}

\subsection{Measurement of  weak identification}

\citet{staiger1994instrumental} proposes to assess the instrument strength using the following $F$-statistic:
\begin{equation*}
    \widehat{F} = \frac{\widehat{\bm{\gamma}}^\top\bm{Z}^\top \bm{Z} \widehat{\bm{\gamma}}}{p \widehat{\sigma}^2_\delta}, 
\end{equation*}
where $\widehat{\bm{\gamma}}$ denotes the  coefficient vector by fitting an OLS regression of the exposure on the instruments, and $\widehat{\sigma}^2_\delta$ is the corresponding residual variance. This statistic provides a test of the joint null hypothesis $\bm{\gamma} = \bm{0}$ in the first-stage regression of 2SLS, and is therefore commonly referred to as the ``first-stage $F$-statistic" \citep{staiger1994instrumental, stock2002testing}.  Under the null hypothesis $\bm{\gamma} = \bm{0}$ and within the weak instrument asymptotics framework (i.e., IV strengths $\{\gamma_j\}_{j=1}^p$ shrink at a $1/\sqrt{n}$ rate \citep{staiger1994instrumental}),  $p\widehat{F}$ converges in distribution to a noncentral chi-squared random variable with $p$ degrees of freedom and noncentrality parameter $\mu^2$ \citep{stock2002testing}.  As suggested by \citet{staiger1994instrumental}, $\widehat{F}<10$ is the rule-of-thumb threshold for weak instruments.

\subsection{Addressing weak identification bias in MR Studies}
Recent developments in IV and MR literature have advanced methodologies to address weak instrument bias. For example, \citet{ye2021debiased} proposes dIVW, a debiased version of the inverse-variance weighted (IVW) estimator, which is robust to many weak IVs using two-sample summary-level data. \citet{xu2023novel} further develops the penalized IVW (pIVW) estimator by using a penalization approach to prevent the denominator of dIVW estimator to be too close to zero. \citet{mikusheva2022inference} defines weak identification in the context of many instruments, where the number of instruments $p$ grows with the sample size $n$, and introduces a jackknifed version of the Anderson-Rubin test statistic \citep{anderson1949estimation} that is robust to weak identification with many instruments and heteroscedasticity in both the exposure and the outcome.  \citet{ye2024genius} proposes GENIUS-MAWII (G-Estimation under No Interaction with Unmeasured Selection leveraging MAny Weak Invalid IVs), which simultaneously addresses the challenges of many weak instruments and widespread horizontal pleiotropy in MR studies.

\section{Population-based  versus Family-based Design}
\label{sec: family-based}
\subsection{Population-based MR design}
Population-based designs (e.g., cohort studies and case-control studies) include unrelated subjects from the target population \citep{szklo1998population, nkomo2006burden, rothman2008modern}. A cohort study is an observational research method where a group of people with a shared characteristic, called a cohort, is followed over time to observe health outcomes or the development of a disease after a specific exposure \citep{szklo1998population, rothman2008modern}. These studies identify groups based on factors like exposure to a risk factor and then compare the outcomes in exposed versus unexposed individuals to determine associations. There are two main types of cohort studies: prospective cohort studies \citep{sedgwick2013prospective}, which follow the group into the future, and retrospective cohort studies \citep{sedgwick2014retrospective}, which look back at historical data.  For example, UK Biobank (UKB) is a large-scale, prospective cohort study that includes over 500,000 participants aged 40-69 at recruitment across the United Kingdom \citep{sudlow2015uk,bycroft2018uk}. In contrast, a case-control study retrospectively compares subjects with a specific outcome (cases) to those without (controls) \citep{schlesselman1982case, breslow1996statistics, rothman2008modern}. Since outcomes are often rare, cases are oversampled and controls are undersampled in case-control studies, resulting in a sample that may not reflect the target population \citep{wan2021matched}.  Nevertheless, logistic regression can still provides valid association estimates on the odds ratio scale in case-control studies \citep{prentice1979logistic}. MR studies leveraging population-based designs benefit from large sample sizes and wide coverage. For example, UKB has genotyped over 500,000 individuals, providing high statistical power to detect modest associations between exposures and outcomes \citep{sudlow2015uk,bycroft2018uk}. However, population-based designs are susceptible to confounding by population stratification, assortative mating, dynastic effects, and selection bias \citep{lawlor2008mendelian, brumpton2020avoiding, sanderson2022mendelian}. To mitigate these biases, researchers often apply methods such as adjusting for principal components, matching, or the use of negative controls \citep{price2006principal, stuart2010matching, lipsitch2010negative, sanderson2021use}.

\subsection{Family-based MR design}
Family-based designs in MR use data from related individuals, typically sibling pairs or parent-offspring trios, to draw causal conclusions within families \citep{davies2019within, brumpton2020avoiding, howe2022within, lapierre2023leveraging, davies2024importance}. By comparing genetically and demographically similar relatives, these designs inherently control for many confounding factors \citep{kong2018nature, howe2022within,davies2024importance}. For example, in a sibling-based MR, one sibling can serve as a control for shared family background \citep{howe2022within}. This within-family comparison helps to eliminate bias due to population stratification, dynastic effects, and assortative mating, which may otherwise confound population-based MR findings \citep{brumpton2020avoiding, howe2022within}. However, family-based study designs typically have smaller sample sizes, limiting statistical power to detect causal relationships \citep{chen2007family, brumpton2020avoiding, howe2022within}. Moreover, these designs require more complex modeling  to properly account for family structures and relatedness among subjects \citep{brumpton2020avoiding, hwang2021integrating, lapierre2023leveraging}. Despite these challenges, family-based MR has proven valuable; for example, within-family analyses have shown that effects of height and BMI on educational attainment, observed in population-based MR, are substantially attenuated when shared familial factors are controlled \citep{brumpton2020avoiding}.

\subsection{Choosing between population and family-based MR designs}

In summary, population-based and family-based MR designs offer complementary advantages and trade-offs, as summarized in Table \ref{tab: population_vs_family}. Population-based MR relies on broad, large-scale samples, offering greater statistical power and generalizability but is more susceptible to bias from population stratification, assortative mating, dynastic effects, and selection bias. Family-based MR inherently accounts for many shared genetic and environmental factors, enhancing the reliability of causal conclusions, but typically involves smaller samples and more complex modeling. In practice, combining insights from both study designs can provide more robust causal findings.

\begin{longtable}{p{2.4cm} p{6.0cm} p{6.0cm}}
\caption{\parbox{\textwidth}{Comparison of population-based and family-based MR designs}}
 \label{tab: population_vs_family} \\
\toprule
\textbf{Feature} & \textbf{Population-based design} & \textbf{Family-based design} \\
\midrule
\endfirsthead

\multicolumn{3}{c}{{\tablename\ \thetable{} -- continued from previous page}} \\
\toprule
\textbf{Feature} & \textbf{Population-based design} & \textbf{Family-based design} \\
\midrule
\endhead

\midrule \multicolumn{3}{r}{{Continued on next page}} \\
\endfoot

\bottomrule
\endlastfoot

Definition & Includes subjects sampled from the target population (e.g., UK Biobank). & Includes genetically related subjects (e.g., siblings or parent-offspring trios). \\
\midrule
Study types & Includes designs such as cohort studies (prospective or retrospective) and case-control studies. & Include designs such as sibling designs and parent-offspring trio designs. \\
\midrule
Key strengths & Generally larger sample size; high statistical power and broader generalizability. & Inherent control for population stratification, dynastic effects, and shared environment. \\
\midrule
Limitations & Susceptible to population stratification, assortative mating, dynastic bias, and selection bias. & Smaller sample sizes; requires more complex modeling to account for family structures and relatedness. \\
\midrule
IV assumption violation risks & More susceptible to violations of the IV independence and exclusion restriction assumptions. & Less susceptible to IV assumption violations due to within-family comparison. \\
\midrule
Bias mitigation & Adjustment for principal components, matching, and negative control outcomes. & Natural control for genetic and environmental confounding within families. \\
\end{longtable}

\section{Individual-level versus Summary-level Data}
\label{sec: ind vs sum}

\subsection{MR methods using individual-level data}

Let $\bm{Y}=(Y_1,\cdots,Y_n)^\top$ be the vector of continuous outcomes of $n$ subjects, $\bm{D}=(D_1,\cdots,D_n)^\top$ be the vector of continuous exposures,  and $\bm{Z}=(Z_{ij})_{n\times p}$ be the matrix of genetic instruments, where $Z_{ij}$ is the genotype of $j$th genetic instrument of subject $i$. \textit{Individual-level data MR} utilizes the dataset $\{\bm{Y}, \bm{D}, \bm{Z}\}$ containing individual-level measurements of outcomes, exposures, and genotypes. When individual-level data are available and all instruments are valid,  two-stage least squares (2SLS) estimator $\widehat{\beta}_{\text{2SLS}}$ defined in equation \eqref{eq: 2SLS} is commonly employed to estimate the causal effect $\beta$  for continuous outcomes under the ALICE model framework \citep{angrist2009mostly, wooldridge2016introductory}. When all genetic instruments satisfy the three core IV assumptions \ref{ass: IV relevance}-\ref{ass: exclusion restriction}, the 2SLS estimator $\widehat{\beta}_{\text{2SLS}}$ is consistent to the causal effect $\beta$ \citep{wooldridge2016introductory}. When some genetic instruments violate one or more of the core IV assumptions, alternative individual-level data MR methods have been developed to estimate the causal effect $\beta$, for example, sisVIVE \citep{kang2016instrumental}, TSHT \citep{guo2018confidence}, MR-GENIUS \citep{tchetgen2021genius}, MR-MiSTERI \citep{liu2023mendelian}, MRSquare \citep{sun2023semiparametric}, GENIUS-MAWII \citep{ye2024genius}, and MR-MAGIC \citep{zhang2025mr}. See Sections \ref{sec: invalid IV} and \ref{sec: weak IV}, as well as \citet{kang2024identification}, for details.

\subsection{MR methods using summary-level data}

In contrast, \textit{summary-level} MR utilizes summary statistics of marginal IV-exposure and IV-outcome associations derived from individual-level data (often not directly accessible) to perform causal inference. Let $\bm{Z}_{\cdot j}=(Z_{1j},\cdots,Z_{nj})^\top$ denote the genotype vector of $j$th genetic instrument, then the marginal estimates of $\gamma_j$ and $\Gamma_j$ are obtained through marginal regressions of $\bm{D}$ and $\bm{Y}$ on $\bm{Z}_{\cdot j}$, respectively:
\begin{align*}
    \widehat{\gamma}_j &= \left(\bm{Z}_{\cdot j}^\top \bm{Z}_{\cdot j}\right)^{-1} \bm{Z}_{\cdot j}^\top \bm{D}, \\
    \widehat{\Gamma}_j &= \left(\bm{Z}_{\cdot j}^\top \bm{Z}_{\cdot j}\right)^{-1} \bm{Z}_{\cdot j}^\top \bm{Y}. 
\end{align*}
Then, the ratio estimator using $j$th genetic instrument is
\begin{equation*}
    \widehat{\beta}_j = \frac{\widehat{\Gamma}_j}{\widehat{\gamma}_j}.
\end{equation*}
When all $p$ genetic instruments are valid, the following inverse-variance weighted (IVW) estimator \citep{burgess2013mendelian} combines ratio estimators from each genetic instrument:
\begin{equation*}
    \widehat{\beta}_{\text{IVW}} = \frac{\sum_{j=1}^p \widehat{\gamma}_j^2 \widehat{\sigma}_{\Gamma,j}^{-2} \widehat{\beta}_j}{\sum_{j=1}^p \widehat{\gamma}_j^2 \widehat{\sigma}_{\Gamma,j}^{-2}},
\end{equation*}
where $\widehat{\sigma}_{\Gamma,j}$ is the standard error of $\widehat{\Gamma}_j$. The IVW estimator upweights genetic instruments with  stronger IV-exposure associations (i.e., larger $\widehat{\gamma}_j^2$) and more precise IV-outcome associations (i.e., smaller $\widehat{\sigma}_{\Gamma,j}^{2}$). In addition, $\widehat{\beta}_{\text{IVW}}$ consistently estimates $\beta$ when all instruments are valid and mutually independent. When some genetic instruments violate the core IV assumptions, the IVW estimator is no longer consistent. To address this, several summary-level data MR methods have been proposed to obtain robust causal effect estimates even in the violation of core IV assumptions \citep{bowden2015mendelian, bowden2016consistent, verbanck2018detection, zhao2019two, burgess2020robust, xu2023novel, yao2024deciphering}. We provide a list of commonly used software implementations with links in Supplementary Materials.

\subsection{Comparison of individual-level and summary-level data in MR}

Compared to summary-level data MR methods, individual-level data MR offers distinct methodological advantages in modeling nonlinear biological relationships and addressing invalid instrument issues. First, individual-level data allow for the characterization of nonlinear relationship among genetic instruments, exposures and outcomes \citep{hall2005nonparametric, veitia2013gene, staley2017semiparametric, guo2022robustness, sulc2022polynomial, sun2023semiparametric}. For example, \citet{guo2022robustness} explicitly models both nonlinear IV-exposure associations and nonlinear violations of assumptions (A2) and (A3), and proposes the Two-Stage Curvature Identification (TSCI) method to identify and estimate the causal effect of interest using individual-level data. In contrast, summary-level data are
calculated using linear or generalized linear models, and thus MR based on summary-level data lacks the capacity to detect nonlinear associations \citep{burgess2024towards}. Second, individual-level data enable more flexible approaches for handling invalid IVs. For example, \citet{sun2023semiparametric} proposes a class of G-estimators for the causal effect in the presence of multiple potentially invalid IVs by leveraging gene-gene interactions, and \citet{liu2023mendelian} proposes novel identification assumptions for  the average treatment effect on the treated (ATT) with a possibly invalid IV.

Conversely, summary-level data MR provides significant practical advantages in genomic data. First, publicly accessible genome-wide association studies (GWAS) summary statistics have become increasingly abundant \citep{yang2012conditional, zhu2016integration, buniello2019nhgri}, overcoming privacy concerns and logistic burdens that often limit access to individual-level genetic data \citep{kaufman2009public,naveed2015privacy, harmanci2016quantification}. Second, GWAS consortia routinely combine data from hundreds of thousands of participants, significantly enhancing statistical power to detect causal relationships \citep{swerdlow2016selecting}. In addition, platforms like MR-Base \citep{hemani2018mr} further streamline analysis by enabling efficient harmonization of exposure and outcome summary statistics across multiple GWAS datasets.

\section{One-sample versus Two-sample Design}
\label{sec: one-sample vs two-sample}
\subsection{Overview and conceptual differences}

To facilitate comparison between one-sample and two-sample MR designs, we adopt the ALICE framework for two independent datasets \citep{angrist1992effect, angrist1995split, zhao2019two}. Let $s\in\{1,2\}$ index two independent samples with sample sizes $n^{(1)}$ and $n^{(2)}$, respectively. Within each sample $s$, let $Y_i^{(s)}$ denote the outcome, $D_i^{(s)}$ denote the exposure, and $\bm{Z}_i^{(s)}=(Z_{i1}^{(s)},\cdots,Z_{ip}^{(s)})^\top\in\mathbb{R}^p$ denote the vector of $p$ genetic instruments for subject $i$. The data $\{Y_i^{(s)}, D_i^{(s)},\bm{Z}_i^{(s)}\}_{i=1}^{n^{(s)}}$ are generated according to:
\begin{align*}
    D_i^{(s)} &= \sum_{j=1}^p  \gamma_j^{(s)} Z_{ij}^{(s)} +  \delta_i^{(s)},  \\
    Y_i^{(s)} &= \beta^{(s)} D_i^{(s)} + \sum_{j=1}^p  \pi_j^{(s)} Z_{ij}^{(s)} + \varepsilon_i^{(s)},  
\end{align*}
with error terms satisfying $\mathbb{E}(\delta_i^{(s)}|\bm{Z}_{i}^{(s)}) = 0$ and $\mathbb{E}(\varepsilon_i^{(s)}|\bm{Z}_{i}^{(s)}) = 0$. We now state the study objectives in one-sample and two-sample MR designs as follows \citep{zhao2019two}:
\begin{itemize}
    \item \textbf{Objective in one-sample MR design}: Given either the individual-level data $\{Y_i^{(s)}, D_i^{(s)}, \bm{Z}_{i}^{(s)}\}_{i=1}^{n^{(s)}}$, or the corresponding summary-level data of IV-exposure and IV-outcome associations from sample $s$, how to  estimate the the  causal effect $\beta^{(s)}$?
     \item \textbf{Objective in two-sample MR design}: Given the individual-level data $\{Y_i^{(1)}, \bm{Z}_{i}^{(1)}\}_{i=1}^{n^{(1)}}$ from the first sample and $\{D_i^{(2)}, \bm{Z}_{i}^{(2)}\}_{i=1}^{n^{(2)}}$ from the second sample (or their corresponding summary-level data), how to  estimate the causal effects $\beta^{(1)}$ and/or $\beta^{(2)}$?
\end{itemize}

As noted in \citet{zhao2019two}, to enable the identification and estimation of the causal effect in two-sample designs, we further need to impose the following assumptions \citep{angrist1992effect, angrist1995split, zhao2019two}:

\begin{assumptionp}{C1}[Homogeneity in parameters]
    $\beta^{(1)}=\beta^{(2)}=\beta$, $\gamma_j^{(1)}=\gamma_j^{(2)}=\gamma_j$, and $\pi_j^{(1)}=\pi_j^{(2)}=\pi_j$ for $j=1,\cdots,p$.  \label{ass: Homogeneity in parameters}
\end{assumptionp}

\begin{assumptionp}{C2}[Homogeneity in the distribution of error terms]
    $(\delta_i^{(1)},\varepsilon_i^{(1)})\overset{d}{=}(\delta_{i'}^{(2)},\varepsilon_{i'}^{(2)})$ for $i\in\{1,\cdots,n^{(1)}\}$ and $i'\in\{1,\cdots,n^{(2)}\}$, where $\overset{d}{=}$  indicates that the random vectors have the same distribution. \label{ass: Homogeneity in error}
\end{assumptionp}

\begin{remark}
    Under assumptions \ref{ass: Homogeneity in parameters} and \ref{ass: Homogeneity in error}, the only source of heterogeneity between the two samples arises from differences in the distribution of instruments \citep{zhao2019two}. In the context of MR, such heterogeneity may reflect differences in genetic ancestry, sampling design, or genotyping platforms, which can lead to differences in allele frequencies or linkage disequilibrium patterns.
\end{remark}

\subsection{ Weak IV biases in one-sample and two-sample MR estimations}

In this section, we focus on the comparison between one-sample and two-sample MR designs, and for simplicity we assume all genetic instruments are valid IVs, i.e., $\pi_j^{(1)}=\pi_2^{(2)}=0$ for $j=1,\cdots,p$. Under assumptions \ref{ass: Homogeneity in parameters} and \ref{ass: Homogeneity in error} and by assuming all genetic instruments are valid IVs, the above ALICE model becomes
\begin{align*}
    D_i^{(s)} &= \sum_{j=1}^p  \gamma_j Z_{ij}^{(s)} +  \delta_i^{(s)},  \\
    Y_i^{(s)} &= \beta D_i^{(s)} + \varepsilon_i^{(s)},  
\end{align*}
and the reduced-form outcome model becomes
\begin{align*}
    Y_i^{(s)} &= \sum_{j=1}^p  \Gamma_j Z_{ij}^{(s)} +  e_i^{(s)},
\end{align*}
where $\Gamma_j=\beta\gamma_j$ and $e_i^{(s)}=\beta \delta_i^{(s)} + \varepsilon_i^{(s)}$. Within sample $s$, let $\bm{Y}^{(s)}=(Y_1^{(s)},\cdots,Y_{n^{(s)}}^{(s)})^\top$ be the vector of outcomes, $\bm{D}^{(s)}=(D_1^{(s)},\cdots,D_{n^{(s)}}^{(s)})^\top$ be the vector of exposures,  and $\bm{Z}^{(s)}=(\bm{Z}_{1}^{(s)},\cdots,\bm{Z}_{n^{(s)}}^{(s)})^\top\in\mathbb{R}^{n^{(s)}\times p}$ be the matrix of genetic instruments. For simplicity, we further assume that genetic instruments are (1) standardized such that $\mathbb{E}(Z_{ij}^{(s)})=0$ and $\text{Var}(Z_{ij}^{(s)})=1$ for $j=1,\cdots,p$ \citep{bulik2015ld}, and (2) mutually independent after LD clumping \citep{purcell2007plink}. We now analyze the  weak instrument biases of one-sample and two-sample 2SLS estimators under this setup.

Let $\widehat{\beta}_{\text{2SLS}}^{(s)}$ denote the one-sample 2SLS estimator using individual-level data from sample $s$. According to \citet{hahn2002notes}, the   weak instrument bias of $\widehat{\beta}_{\text{2SLS}}^{(s)}$ can be approximated as follows:
\begin{equation*}
    \mathbb{E}\left(\widehat{\beta}_{\text{2SLS}}^{(s)}\right) - \beta \approx \frac{ \sigma_{\delta,\varepsilon}}{{n^{(s)}\|\bm{\gamma}\|^2}/{p}+  \sigma_\delta^2},
\end{equation*}
where $\sigma_{\delta,\varepsilon}$ is the covariance between $\bm{\delta}^{(s)}$ and $\bm{\varepsilon}^{(s)}$, $\sigma_\delta^2$ is the variance of $\bm{\delta}^{(s)}$, and $\|\bm{\gamma}\|^2=\sum_{j=1}^p\gamma_j^2$. We also provide the derivation of this approximate bias in Supplementary Materials. On the other hand, the approxmate bias of ordinary least square (OLS) estimator $\widehat{\beta}_{\text{OLS}}^{(s)}$ using sample $s$ is
\begin{equation*}
    \mathbb{E}\left(\widehat{\beta}_{\text{OLS}}^{(s)}\right) - \beta \approx \frac{ \sigma_{\delta,\varepsilon}}{\|\bm{\gamma}\|^2+  \sigma_\delta^2}.
\end{equation*}
With weak instruments, both the one-sample 2SLS and OLS estimators are biased when the error terms in the exposure and outcome models are correlated, i.e., $\sigma_{\delta,\varepsilon}\neq 0$. Importantly, the direction of the bias for $\widehat{\beta}_{\text{2SLS}}^{(s)}$ is the same as that for $\widehat{\beta}_{\text{OLS}}^{(s)}$, implying that the one-sample 2SLS estimator tends to be biased towards the OLS estimator with weak instruments.

In the two-sample design, the IV–exposure associations are first estimated in the first sample and then used to construct fitted exposures in the second sample. The causal effect is subsequently estimated by regression the outcome on these fitted exposures in the second sample. This estimation strategy is also known as the Split-Sample Instrumental Variable (SSIV) estimation \citep{angrist1995split}. Under our setting, the two-sample 2SLS estimator, denoted as $\widehat{\beta}_{\text{SSIV}}$, has the following approximation \citep{angrist1995split}:
\begin{equation*}
    \mathbb{E}\left(\widehat{\beta}_{\text{SSIV}}\right) \approx \beta \times \frac{ \|\bm{\gamma}\|^2}{ \|\bm{\gamma}\|^2 + {p\sigma_\delta^2}/{n^{(1)}}}.
\end{equation*}
This expression shows that the two‐sample 2SLS estimator is attenuated toward zero by a factor that depends on the first‐stage sample size $n^{(1)}$, the number of IVs $p$, the IV strengths, and the variance of the error in exposure model $\sigma_{\delta}^2$. Unlike in the one-sample setting, where weak instruments bias the 2SLS estimator toward the confounded OLS estimate, the two-sample 2SLS estimator is biased toward zero when instruments are weak.

Then, we consider the case where only marginal association estimates and their standard errors are available from a single sample. Specifically, in sample $s$, the marginal estimates of IV-exposure and IV-outcome associations of $j$th genetic instrument are $\widehat{\gamma}_j^{(s)} =  \left((\bm{Z}_{\cdot j}^{(s)})^\top \bm{Z}_{\cdot j}^{(s)}\right)^{-1} (\bm{Z}^{(s)}_{\cdot j})^\top \bm{D}^{(s)}$ and $\widehat{\Gamma}_j^{(s)} =  \left((\bm{Z}_{\cdot j}^{(s)})^\top \bm{Z}_{\cdot j}^{(s)}\right)^{-1} (\bm{Z}^{(s)}_{\cdot j})^\top \bm{Y}^{(s)}$, respectively. Let $\widehat{\sigma}^{(s)}_{\Gamma,j}$ denote the standard error of $\widehat{\Gamma}_j^{(s)}$. Then, the one-sample IVW estimator for the causal effect $\beta$ using summary statistics from sample $s$ is given by
\begin{equation*}
    \widehat{\beta}_{\text{IVW}}^{(s)} = \frac{\sum_{j=1}^p \widehat{\Gamma}_j^{(s)} \widehat{\gamma}_j^{(s)} / (\widehat{\sigma}^{(s)}_{\Gamma,j})^2}{ \sum_{j=1}^p (\widehat{\gamma}_j^{(s)})^2 / (\widehat{\sigma}^{(s)}_{\Gamma,j})^2},
\end{equation*}
and the bias of $\widehat{\beta}_{\text{IVW}}^{(s)}$ can be approximated as
\begin{equation*}
    \mathbb{E}\left(\widehat{\beta}_{\text{IVW}}^{(s)}\right) - \beta \approx \frac{\sigma_{\delta,\varepsilon}}{n^{(s)}}\times\frac{\sum_{j=1}^p 1/(\widehat{\sigma}^{(s)}_{\Gamma,j})^{2}}{\sum_{j=1}^p (\gamma_j^2+(\sigma_{\gamma,j}^{(s)})^{2})/(\widehat{\sigma}^{(s)}_{\Gamma,j})^{2}},
\end{equation*}
where $(\sigma_{\gamma,j}^{(s)})^{2}$ is the variance of $\widehat{\gamma}_j^{(s)}$. As with the one-sample 2SLS estimator, this weak instrument bias arises due to the correlation between error terms in the exposure and outcome models, and tends toward the OLS estimator.

Finally, we consider the two-sample IVW estimator that combines the the IV–exposure association estimates $\{\widehat{\gamma}_j^{(1)}\}_{j=1}^p$ from the first sample and the IV–outcome association estimates $\{\widehat{\Gamma}_j^{(2)}\}_{j=1}^p$ from the second sample \citep{burgess2013mendelian, bowden2017framework}, which is given by 
\begin{equation*}
    \widehat{\beta}_{\text{IVW}}^{(1,2)} = \frac{\sum_{j=1}^p \widehat{\Gamma}_j^{(2)} \widehat{\gamma}_j^{(1)} / (\widehat{\sigma}^{(2)}_{\Gamma,j})^2}{ \sum_{j=1}^p (\widehat{\gamma}_j^{(1)})^2 / (\widehat{\sigma}^{(2)}_{\Gamma,j})^2},
\end{equation*}
and the expectation of the two-sample IVW estimator can be approximated as follows \citep{zhao2020statistical, ye2021debiased}:
\begin{equation*}
    \mathbb{E}\left(\widehat{\beta}_{\text{IVW}}^{(1,2)}\right) \approx \beta \times \frac{\sum_{j=1}^p \gamma_j^2 / (\widehat{\sigma}^{(2)}_{\Gamma,j})^2}{\sum_{j=1}^p (\gamma_j^2+(\sigma^{(1)}_{\gamma,j})^2) / (\widehat{\sigma}^{(2)}_{\Gamma,j})^2  }.
\end{equation*}
This reveals that the two-sample IVW estimator is biased toward zero with weak instruments, similar to the two-sample 2SLS estimator.

\begin{conclusion}
In the presence of weak IVs, one-sample MR estimators tends to be biased towards the confounded OLS estimator, whereas two-sample MR estimators tends to be biased towards zero.
\end{conclusion}

\begin{remark}
    To handle the weak IV bias in two-sample summary-level data MR analysis, \citet{xu2023novel} proposes a novel penalized inverse-variance weighted (pIVW) estimator that adjusts the IVW estimator through a penalized likelihood approach. 
\end{remark}

\subsection{Advantages, limitations, and recommendations for practice}

In summary, both the 2SLS estimator with individual-level data and the IVW estimator with summary-level data in one-sample study designs are biased toward the OLS estimator with weak instruments. In contrast, both 2SLS and IVW estimators in two-sample study designs tends to be biased toward zero. In addition, in two-sample study design, the identification of the causal effect and interpretation of the estimate require assumptions \ref{ass: Homogeneity in parameters} and \ref{ass: Homogeneity in error} on parameters and error terms in addition to core IV assumptions. Neither study design is universally superior; the choice between one-sample and two-sample study design depends on data availability (single versus two independent datasets), interpretability (population-specific versus generalizable estimates), and bias trade-offs (weak IV bias toward confounded OLS estimates versus toward zero).

\section{Selecting Genetic IVs using the Anna Karenina Principle}
\label{sec: IV selection}

A critical step in MR analysis is the selection of appropriate genetic variants to serve as instruments. In practice, variants are often chosen based on their strength of association with the exposure, typically using GWAS summary statistics. Thresholds such as $p < 5 \times 10^{-8}$ (genome-wide significance) or $p < 1 \times 10^{-6}$ are commonly applied, although the precise cut-off varies across studies \citep{panagiotou2012should, swerdlow2016selecting, kanai2016empirical, sanderson2022mendelian}.  

The more difficult challenge lies in distinguishing valid from invalid instruments when the core assumptions may be violated. Different MR methods address this issue by introducing additional identification assumptions. For example, MR-PRESSO \citep{verbanck2018detection} identifies valid IVs under the \emph{majority rule}.  

Motivated by the Anna Karenina Principle (AKP) \citep{yao2024deciphering}, which states that "all happy families are alike, but every unhappy family is unhappy in its own way," we view valid instruments as a coherent group that share the same properties, while each invalid instrument may fail validity in a distinct manner. Building on this intuition, MR-SPI adopts the \emph{plurality rule assumption} \ref{ass: plurality}, which requires only that the largest group of instruments corresponds to the valid set, even if valid instruments do not form a majority.  

The SPI procedure operationalizes this idea by identifying the largest cluster of variants that conform to the IV assumptions, while allowing for heterogeneous violations among the rest. This strategy reflects the AKP perspective: validity is uniform, but invalidity can be idiosyncratic. By leveraging this asymmetry, MR-SPI provides a principled and robust approach to selecting instruments from GWAS summary data. More specifically, the selection steps are as follows (Figure \ref{fig: SPI selection}):

\begin{figure}[!thb]
    \centering
    \includegraphics[width=0.8\linewidth]{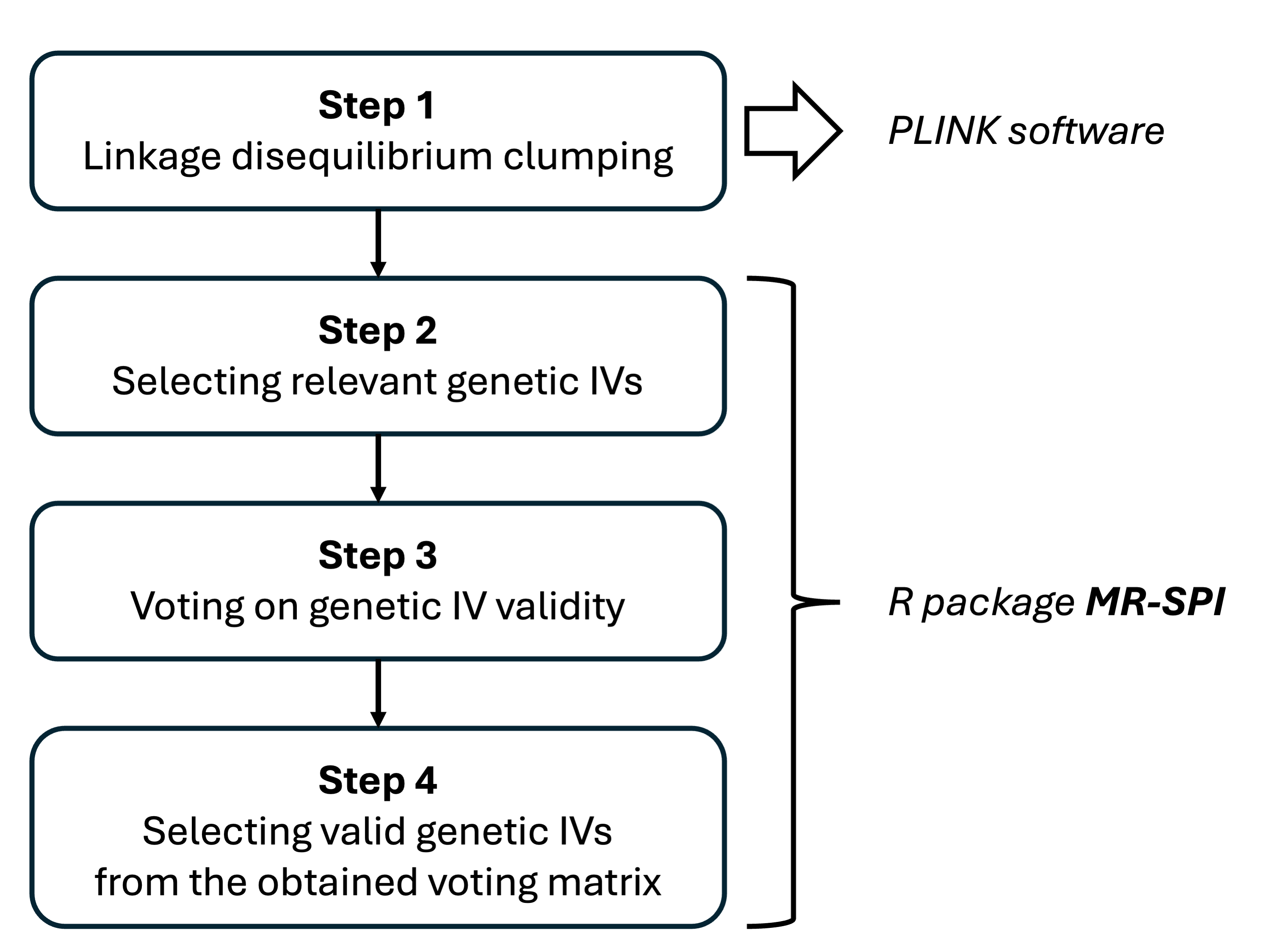}
    \caption{IV selection procedure of MR-SPI.}
    \label{fig: SPI selection}
\end{figure}

\begin{enumerate}[topsep=0pt,itemsep=-1ex,partopsep=1ex,parsep=1ex]
    \item[Step 1.] \textbf{Linkage disequilibrium (LD) clumping.} Obtain a set of approximately independent genetic variants by removing those in high LD with one another using PLINK \citep{purcell2007plink}. 
    \item[Step 2.] \textbf{Selecting relevant genetic IVs.} From the LD-clumped genetic variants, retain those showing strong associations with the exposure using a specific $p$-value thresholding (e.g., $5\times 10^{-8}$).
    \item[Step 3.] \textbf{Voting on genetic IV validity.} 
    For each relevant genetic IV $j$, compute the ratio estimate $\widehat{\beta}_j=\widehat{\Gamma}_j/\widehat{\gamma}_j$. Then, treating $\widehat{\beta}_j$ as the true causal effect, for every other relevant genetic IV $k$, calculate its estimated degrees of violation of assumptions \ref{ass: IV independence} and \ref{ass: exclusion restriction}, defined as $\widehat{\pi}_j^{[k]}=\widehat{\Gamma}_k-\widehat{\beta}_j\widehat{\gamma}_k$. A small $|\widehat{\pi}_j^{[k]}|$ suggests that $k$th genetic IV is also likely to be a valid IV by assuming $j$th genetic IV is valid, and thus $k$th genetic IV ``votes for" $j$th genetic IV to be valid.
    \item [Step 4.] \textbf{Selecting valid genetic IVs from the obtained voting matrix.} Select valid genetic IVs by finding the maximum clique of the voting matrix that encodes whether two relevant IVs mutually vote for each other to be valid.
\end{enumerate}
In practice, Step 1 can be performed via PLINK by applying a predefined pairwise correlation threshold (e.g., $r^2<0.01$) within a specified genomic window (e.g., 1Mb). Steps 2-4 can be implemented using the R package \textbf{MR-SPI} (\url{https://github.com/MinhaoYaooo/MR-SPI}). The instruments selected through these steps can then be used for downstream MR analyses (e.g., IVW method). If the plurality rule assumption \ref{ass: plurality} holds, this procedure can help exclude invalid IVs and enhance the robustness of MR findings.

Several considerations are important when selecting genetic instruments. Ideally, the sample used for IV selection should be independent of the samples used for estimating causal effect to minimize the winner's curse \citep{jiang2023empirical, ma2023breaking}. For example, \citet{zhao2019powerful} proposes a three-sample MR design to eliminate the bias due to the winner's curse.  It is also advisable to use external resources, such as PhenoScanner \citep{kamat2019phenoscanner}, to screen candidate genetic IVs for associations with potential confounders or secondary traits, thereby improving instrument validity. Careful attention to these issues enhances the reliability and reproducibility of MR findings.

\section{Applications in Real Datasets}
\label{sec: real data}

\subsection{Application 1: assessing the causal effect of body mass index on diastolic blood pressure using one-sample individual-level data from UK Biobank}

In this section, we apply several one-sample MR methods to assess the causal effect of body mass index (BMI) on diastolic blood pressure (DBP). This analysis utilizes data from the UK Biobank (UKB) cohort study, a biomedical database comprising genetic and phenotypic information from approximately 500,000 UK participants \citep{sudlow2015uk,bycroft2018uk}. Participants who reported using anti-hypertensive medication or had missing data were excluded, resulting in a final sample of 254,502 individuals. Following \citet{sun2023semiparametric}, we selected the top 10 independent single-nucleotide polymorphisms (SNPs) most strongly associated with BMI after applying linkage disequilibrium (LD) clumping with $r^2<0.01$. These SNPs are rs1558902, rs6567160,  rs543874, rs13021737, rs10182181, rs2207139, rs11030104, rs10938397, rs13107325, and rs3810291.

\begin{figure}[!htb]
    \centering
    \includegraphics[width=0.6\linewidth]{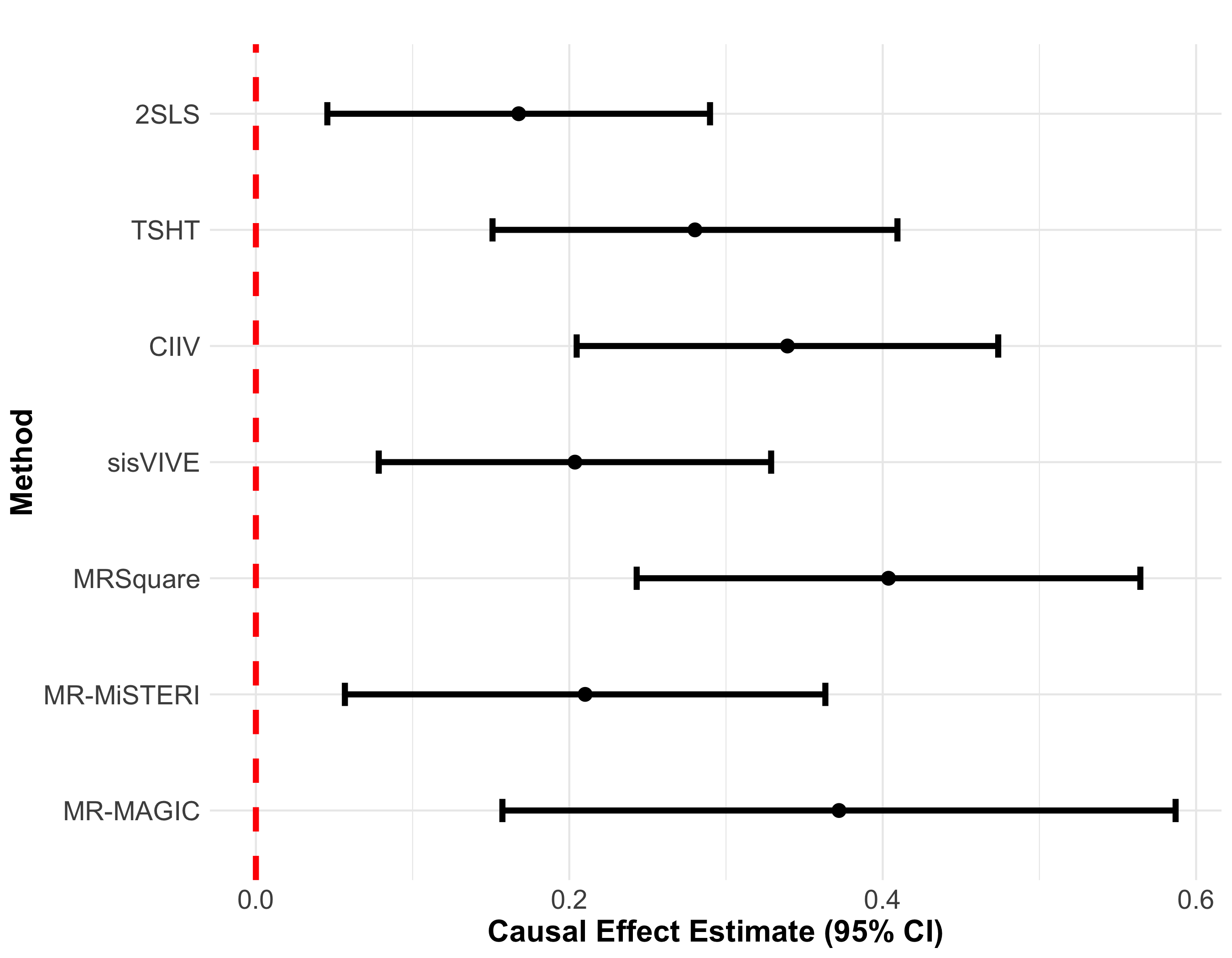}
    \caption{Point estimates and 95\% confidence intervals for the causal effect of body mass index (BMI) on diastolic blood pressure (DBP) in the UK Biobank data, obtained using different one-sample MR methods.}
    \label{fig: UKB results}
\end{figure}

We compare the following methods for estimating the causal effect $\beta$: (1) two-stage least squares (2SLS) \citep{wooldridge2016introductory}; (2) Two-Stage Hard Thresholding (TSHT) \citep{guo2018confidence}; (3) Confidence Interval method for Instrumental Variable (CIIV) \citep{windmeijer2021confidence}; (4) Some Invalid Some Valid IV Estimator (sisVIVE) \citep{kang2016instrumental}; (5) MRSquare \citep{sun2023semiparametric}; MR Mixed-Scale Treatment Effect Robust Identification (MR-MiSTERI) \citep{liu2023mendelian}; and MR with MAny weak Genetic Interactions for Causality (MR-MAGIC) \citep{zhang2025mr}. For sisVIVE, we choose the tuning parameter via 10-fold cross-validation. For MRSquare, we set the minimum number of valid instruments to be $6$. The code for this application is provided in ``DataExamples.R" within the Supplementary Files. The results are summarized in Figure \ref{fig: UKB results}. 

From Figure \ref{fig: UKB results}, all methods suggest a positive causal effect of BMI on DBP, with point estimates ranging from 0.1677 to 0.4037. The 2SLS method yields the smallest estimate ($\widehat{\beta}_{\text{2SLS}} = 0.1677$; 95\% CI: 0.0456–0.2898), likely due to the inclusion of invalid instruments in the analysis. Unlike 2SLS, which assumes all instruments are valid, the other methods account for potential invalid IVs in various ways. Notably, TSHT, CIIV, and sisVIVE implement procedures to select valid instruments from  candidate ones. In this application, TSHT identifies two invalid IVs (rs10182181 and rs13107325), CIIV identifies three (rs10182181, rs13107325, and rs3810291), and sisVIVE identifies one (rs13107325). Instrument rs13107325 is consistently identified as an invalid IV by all three methods.

\subsection{Application 2: performing xMR analysis to identify plasma proteins associated with the risk of Alzheimer's disease using two-sample summary-level data}

The increasingly available large-scale multi-omics data (e.g., epigenomics, transcriptomics, proteomics, and metabolomics data) enable us to perform omics MR (xMR, firstly coined in \citet{yao2024deciphering}), a methodology within the \textit{causal genomics} field, to detect putative causal omics biomarkers for complex traits and diseases, thereby uncovering the underlying causal mechanisms. For a detailed, step‑by‑step tutorial on implementing commonly used xMR methods, please refer to \citet{yao2025introduction}.

In this section, we apply several two-sample MR methods to perform xMR analysis, aiming to identify putative causal plasma proteins associated with the risk of Alzheimer's disease. For the exposure, we use UK Biobank Pharma Proteomics Project (UKB‑PPP) summary statistics on 1,463 plasma proteins measured in 54,306 individuals \citep{sun2023plasma}. For the outcome, we use summary statistics from a meta-analysis of GWASs for clinically diagnosed AD and AD by proxy, comprising 455,258 samples in total \citep{jansen2019genome}. Genetic instruments for each protein are selected by applying a Bonferroni‐corrected threshold of $p\text{-value}<3.40\times10^{-11}$, followed by LD clumping at threshold $r^2<0.01$, as described in \citet{sun2023plasma}. We compare the following two-sample MR methods in this application: (1) inverse-variance weighted (IVW) method \citep{bowden2017framework}; (2) MR-Egger regression \citep{bowden2015mendelian}; (3) MR using
the Robust Adjusted Profile Score (MR-RAPS) \citep{zhao2020statistical}; (4) MR Pleiotropy RESidual Sum
and Outlier test (MR-PRESSO) \citep{verbanck2018detection}; (5) the weighted median method \citep{bowden2016consistent}; (6) the mode-based estimation \citep{hartwig2017robust}; (7) MRMix \citep{qi2019mendelian}; (8) the contamination mixture method \citep{burgess2020robust}; and (9) MR with valid IV Selection and
Post-selection Inference (MR-SPI) \citep{yao2024deciphering}. The code for this application can be found in ``DataExamples.R" within the Supplementary Files.

Proteins identified to be significantly associated with Alzheimer’s disease after Bonferroni correction \citep{bland1995multiple} are summarized in Figure S1 in Supplementary Materials. In Figure S1(A), we report the number of significant plasma proteins detected by each method, which ranges from 0 (MR‑PRESSO) to 14 (MRMix). MR‑PRESSO detects no significant proteins, likely because the small number of candidate IVs per protein limits its power to perform the outlier test and detect invalid instruments. In Figure S1(B), we list the 11 plasma proteins identified by at least two methods. Notably, the seven proteins identified by MR‑SPI correspond to the top seven proteins ranked by the number of supporting methods. For a more detailed discussion of these seven proteins including gene ontology analysis and AlphaFold3‑based structural prediction \citep{abramson2024accurate}, please refer to \citet{yao2024deciphering}.

\section{Future directions}
\label{sec: future}

\subsection{Binary and survival outcomes}

In epidemiological research, the binary and survival outcomes are common, yet current MR methods predominantly focus on continuous outcomes. For one-sample MR with individual-level data, establishing identification conditions for MR analysis with binary or survival outcomes is essential, particularly in the presence of invalid instruments \citep{clarke2010identification, deng2023bayesian, liu2025quasi, bu2025semiparametric}. For two-sample MR with summary-level data, the current standard practice is to directly apply existing two-sample MR methods under the ALICE model framework for continuous outcomes to analyze summary statistics of binary or survival outcomes, while the interpretation of the causal effect estimate obtained by this direct application is unclear and requires further justification \citep{zhao2019two}.

\subsection{Longitudinal studies}

A longitudinal study is a research design involving repeated measures of the same variables over prolonged periods of time, widely used in epidemiology and social science to track trends and establish causal relationships. However, longitudinal studies aimed at estimating causal effects may be subject to bias arising from unmeasured confounding and/or time-varying confounding variables. MR analysis can mitigate such unmeasured and time-varying confounding bias by leveraging genetic variations as IVs, though it typically relies on data measured at a single time point. Developing MR methods tailored for longitudinal studies holds promise for estimating time-varying causal effects, thereby providing novel biological insights into lifetime health trajectories \citep{labrecque2019interpretation, morris2022interpretation, sanderson2022estimation}.

\subsection{Multivariate MR}

Multivariate MR extends the classical MR framework to estimate the causal effects of multiple exposures on an outcome simultaneously \citep{burgess2015multivariable, sanderson2019examination}. This approach is particularly valuable when exposures are biologically correlated, and has been applied to disentangle complex causal relationships, for example, identifying metabolite biomarkers for age-related macular degeneration \citep{zuber2020selecting}. Recent studies have begun to address challenges arising from invalid instruments in multivariate MR framework \citep{liang2022selecting, lin2023robust, chan2024novel}. However,  methods robust to the violation of core IV assumptions when handling high-dimensional exposures (e.g., omics biomarkers) are still lacking.

\section*{Acknowledgements}

This research has been conducted using the UK Biobank Resource under application number 52008. The authors express sincere thanks to Dr. Paul Albert for initiating this tutorial and for his constructive feedback during its development.

\bibliography{lit.bib}
\bibliographystyle{apalike}

\end{document}